\documentclass{jfm}
\usepackage{graphicx}
\usepackage{epstopdf, epsfig}

\usepackage{amsmath}
\usepackage{amstext}
\usepackage{amssymb} 
\graphicspath{{./figures/}}
\usepackage{textcomp}
\usepackage{float}
\usepackage{enumitem}

\usepackage{xcolor} 

\shorttitle{Simulation of oscillating
boundary layers on adiabatic slopes}
\shortauthor{B. E. Kaiser, L. J. Pratt and J. Callies}

\title{Direct numerical simulation of low Reynolds number oscillating 
boundary layers on adiabatic slopes}

\author{Bryan E. Kaiser\aff{1}
  \corresp{\email{bkaiser@lanl.gov}},
  Lawrence J. Pratt\aff{2}
 \and J{\"o}rn Callies\aff{3}}

\affiliation{\aff{1}Los Alamos National Laboratory, Los Alamos, NM,
USA
\aff{2}Woods Hole Oceanographic Institution, Woods Hole, MA, USA,
\aff{3}Caltech, Pasadena, CA, USA}

\begin{document}

\maketitle

\begin{abstract}
We investigate the instabilities and transition mechanisms of Boussinesq 
stratified boundary layers on sloping boundaries when subjected to oscillatory 
body forcing parallel to the slope. Such conditions are typical of the boundary layers generated by low wavenumber internal tides sloshing up and down adiabatic abyssal slopes in 
the absence of mean flows, high wavenumber internal tides, and resonant tide-bathymetry interactions. We examine flows within a region of non-dimensional parameter space typical of the mid- to low-latitude oceanic $M_2$ tides on hydraulically smooth abyssal slopes by 
direct numerical simulation. 
We find that at low Reynolds numbers transition-to-turbulence pathways arise from both shear and gravitational instabilities, and we find that the boundary layers are stabilized by increased outer boundary layer stratification during the downslope oscillation phase. However, if rotation is significant (low slope Burger numbers) we find that boundary layer turbulence is sustained throughout the oscillation period, resembling Stokes-Ekman layer turbulence. 
Our results suggest that oscillating boundary layers on smooth abyssal slopes created by
low wavenumber $M_2$ tides do not cause significant irreversible 
turbulent buoyancy flux (mixing) and that flat-bottom dissipation rate models derived from the tide amplitude are accurate within an order of magnitude.
\end{abstract}


\section{Introduction}
Irreversible buoyancy flux convergence within 
oscillating stratified boundary layers on sloping bathymetry in the abyss 
may be a significant mechanism driving the deep branch of the 
global overturning circulation
(\citet{Ferrari16}). The boundary layers, combined with other 
bottom-intensified sources of turbulence, such as the breaking of internal waves, 
contribute to observed patterns of intense irreversible buoyancy flux convergence (\citet{Polzin96b}). How unstable are 
these boundary layers on typical abyssal slopes, and what are the instability mechanisms?
In this article, we employ theory and direct numerical simulations to investigate 
the pathways between laminar, transitional, and turbulent states of boundary layers that are forced by the $M_2$ barotropic tide and occur 
on hydraulically smooth abyssal slopes in the absence of forcing by high-wavenumber internal waves, 
mean flows, far-field turbulence on larger scales, and resonant tidal-bathymetric interaction.

 \begin{figure}
 \centering
 \includegraphics[width=4in]{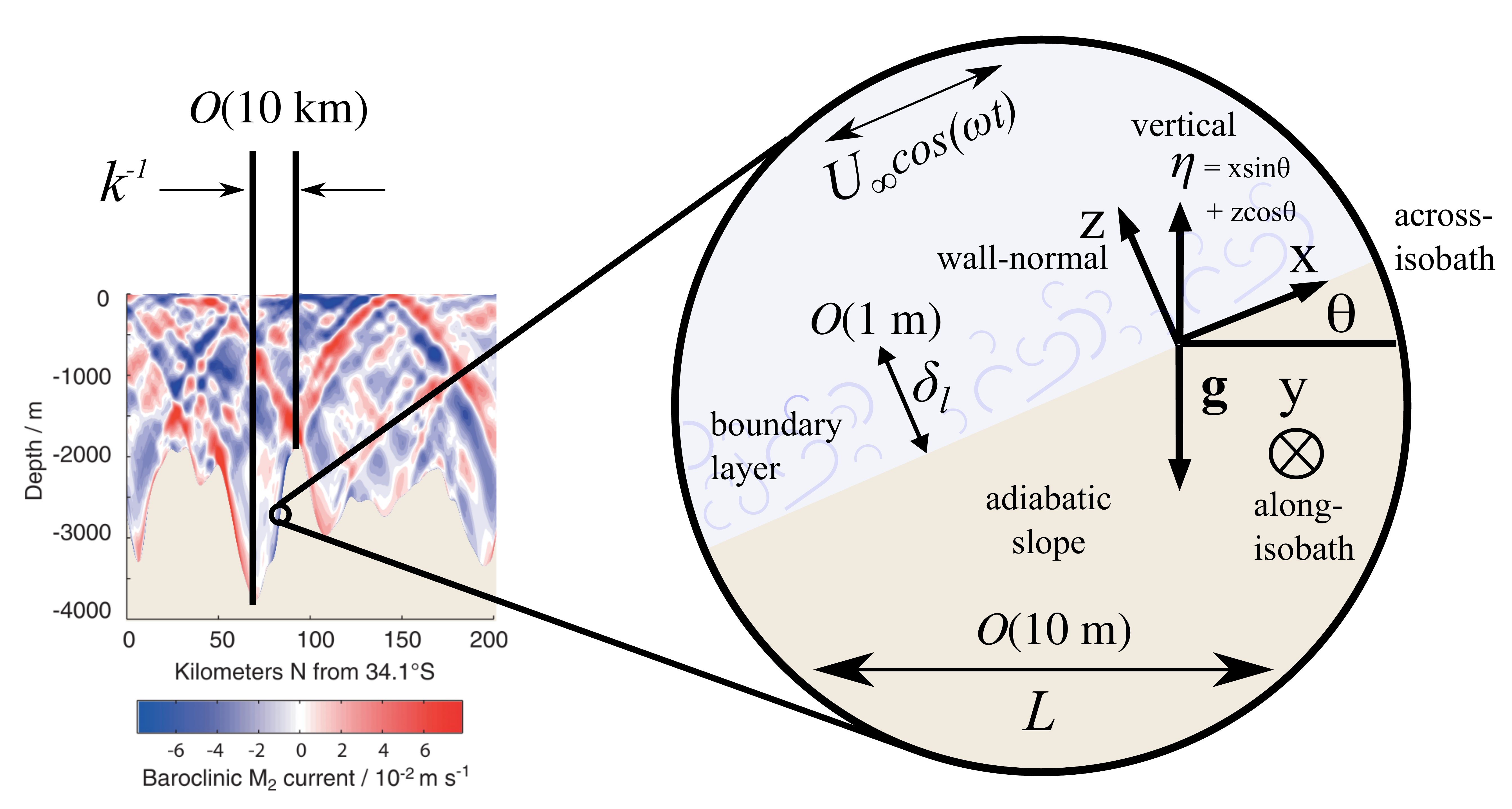}
 \caption{Illustration of NRBTBL length scales and geometry.}
 \label{fig:domain}
  \text{The M$_2$ tide velocity contour plot on the left is 
 from \citet{Zilberman09}. 
 }
 \end{figure}

The boundary layers are formed as momentum and buoyancy are diffused 
by no-slip and adiabatic boundary conditions on sloping bathymetry as 
low-wavenumber internal waves 
heave isopycnals (constant-density contours) up and down 
slopes. Figure \ref{fig:domain} illustrates the scale separations between 
the horizontal length scale of baroclinic tide generation, 
$k^{-1}$, (\citet{Garrett07}), 
the excursion length scale of the tide, $L$, and 
the largest boundary layer length scale, $\delta_l$. 
The excursion length scale, $L$, 
is the characteristic scale of the across-isobath
distance between the trajectory extrema in the inviscid problem, defined 
\begin{equation}
 L\sim{}\frac{U_\infty}{\omega},
 \label{eq:excursion_length}
\end{equation}
where $U_\infty$ is the barotropic tide amplitude projected 
onto the across-slope tangential coordinate ($x$) 
and $\omega$ is the tide frequency.
We assume that the slope curvature can be well approximated 
as constant on scales of the order of the excursion length, $L$. 
Thus the boundary layers investigated in this Article 
apply to boundary layer flows on 
hydraulically smooth slopes where 
the excursion parameter
\begin{equation}
 \mathcal{E}=kL,
 \label{eq:excursion_number}
\end{equation}
is small, $\mathcal{E}\ll1$. 
The excursion parameter $\mathcal{E}$ is  
the 
ratio of net fluid advection by
the barotropic tide 
to the topographic length. 
The baroclinic response to the barotropic forcing is 
highly nonlinear for large excursion parameter flows 
$\mathcal{E}$ (\citet{Bell75a}, \citet{Bell75b}, \citet{Garrett07}, \citet{Sarkar17}).
Here, we investigate the dynamics on scales at or smaller than 
the excursion length, thus we assume that the 
baroclinic tide (a.k.a. internal waves) generated by bathymetric features with horizontal 
length scales of $k^{-1}$ can be locally approximated as 
irrotational over $L$. Therefore we assume that the baroclinic tide
can be modeled as an
across-isobath oscillating body force on the scale of the 
excursion 
length.

Our objectives are 1) to determine if these boundary layers are laminar, transitional, 
 intermittent, or 
 fully turbulent for typical abyssal ocean non-dimensional parameters associated 
 with the $M_2$ tide, 2) to investigate the transition mechanisms, and
 3) to test back-of-the-envelope estimation for
 barotropic tide dissipation rates at the seafloor.
 This Article is organized as follows. In \textit{Problem formulation}
we discuss the relevant governing equations
and non-dimensional parameters.
 In \textit{Linear Flow Solutions} we investigate analytical solutions 
 for the laminar flows to estimate the necessary conditions for boundary layer gravitational 
 instabilities.
In \textit{Nonlinear Flow Solutions} we analyze the stability of simulated boundary layers, and in \textit{Conclusions} we summarize the observed transition mechanisms 
and drag coefficients.

\section{Problem formulation}
The flows examined in this study subject to a body force 
in the across-isobath, or streamwise, direction (the $x$ direction in Figure \ref{fig:domain}), 
\begin{equation}
 F_d(t_\mathrm{d})
 =-\Real[A_\mathrm{d}\text{ie}^{\text{i}\omega t_\mathrm{d}}],
 \label{eq:forcing}
\end{equation}
where $A_\mathrm{d}$ is the dimensional amplitude of the 
pressure gradient $\partial_x\tilde{P}_\mathrm{d}$ and $t_\mathrm{d}$ is dimensional time.

Several geometric and physical approximations are invoked for 
the sake of tractibility and conceptual simplicity.
The flow is approximated as Boussinesq: the density 
variations are small enough that the incompressibility condition is justified, and
Joule heating (increases of internal energy 
due to the viscous dissipation of mechanical energy) 
is neglected. 
We idealize
abyssal buoyancy 
as a linear function of temperature alone.

A Cartesian coordinate system, rotated $\theta$ radians
counterclockwise above the horizontal (Figure \ref{fig:domain}), 
was chosen for analytical
convenience. 
The $z$ coordinate is the wall-normal (or transverse) 
coordinate, which is at angle 
$\theta$ from the vertical coordinate
(the coordinate anti-parallel to the gravity). To distinguish between the 
slope-normal and vertical coordinates, the vertical coordinate (and vertical 
velocities, fluxes, etc) in the direction normal to Earth's surface will 
be denoted as $\eta$, such that $\eta=x\sin\theta+z\cos\theta$, shown in 
Figure \ref{fig:domain}.

The non-dimensional Boussinesq governing equations for 
conservation of mass, momentum, and thermodynamic energy for the flow are
\begin{flalign}
  \partial_x\tilde{u}&+\partial_y\tilde{v}+\partial_z\tilde{w}=0,
  \label{eq:continuity}\\
  \mathrm{d}_t\tilde{{u}}
  &=\frac{1}{\textit{Ro}}\tilde{v}-\partial_x\tilde{p}+
  \frac{1}{\Rey_L}\big(\partial_{xx}+\partial_{yy}+\partial_{zz}\big)\tilde{u}+\textit{C}^2\tilde{b}
  +F(t)
  \label{eq:xmomentum}\\
   \mathrm{d}_t\tilde{{v}}
  &=-\frac{1}{\textit{Ro}}\tilde{u}-\partial_y\tilde{p}+
  \frac{1}{\Rey_L}\big(\partial_{xx}+\partial_{yy}+\partial_{zz}\big)\tilde{v},\\
  \mathrm{d}_t\tilde{{w}}
  &=-\partial_z\tilde{p}+
  \frac{1}{\Rey_L}\big(\partial_{xx}+\partial_{yy}+\partial_{zz}\big)\tilde{w}+\textit{C}^2\tilde{b}\cot\theta,\\
  \mathrm{d}_t\tilde{{b}}
  &=
    \frac{1}{\Pran\hspace{0.3mm}\Rey_L}
    \big(\partial_{xx}+\partial_{yy}+\partial_{zz}\big)\tilde{b},
    \label{eq:buoyancy_nondim}
\end{flalign}
where 
$\mathrm{d}_t=\partial_t+\tilde{\mathbf{u}}\cdot\nabla$ and 
$F(t)=F_\mathrm{d}(t_\mathrm{d})/(U_\infty\omega)=A\sin(t)$.
The variables 
are non-dimensionalized as follows (subscript ``d'' denoting 
dimensional variables): 
\begin{flalign}
\mathbf{x}={\mathbf{x}_\text{d}}/{L}, \qquad
\tilde{\mathbf{u}}={\tilde{\mathbf{u}}_\text{d}}/{U_{\scalebox{.9}{$\scriptscriptstyle \infty$}}}, \qquad
t={\omega}{t_\text{d}}, \qquad \tilde{p}={\tilde{p}_\text{d}}/{U_{\scalebox{.9}{$\scriptscriptstyle \infty$}}^2}, \qquad
\tilde{b}={\tilde{b}_\text{d}}/{(LN^2\sin\theta)}, 
 \label{eq:non_dim_def}
\end{flalign}
where the reference density $\rho_0$ is absorbed into the 
mechanical pressure $p_\text{d}$ such that it has units J$\hspace{0.5mm}$kg$^{-1}$.
$N^2$ is the square of the buoyancy frequency (the natural frequency associated with the restoring 
force of stratification). The buoyancy is defined as the acceleration associated 
with density anomalies, 
$b_\mathrm{d}=g(\rho_0-\rho)/\rho_0$, where $g$ is the (constant) gravity and 
$\rho$ is the density.

Despite the assumptions and idealizations listed above, 
the dynamical 
parameter space is vast. 
The relevant non-dimensional ratios are
the Prandtl number $\Pran$, the slope Rossby number $\textit{Ro}$, 
the slope frequency ratio $\textit{C}$,
and Stokes layer Reynolds number $\Rey$. In this study, the 
Stokes layer Reynolds number is referred to in the analysis of 
the flow instead of the excursion 
length Reynolds number 
\begin{equation}
   \Rey_L=\frac{U_\infty{}L}{\nu}=\frac{U_\infty^2}{\nu\omega},
\end{equation}
because the Stokes layer Reynolds number is common in literature 
regarding oscillating boundary layers.
The Prandtl and Stokes layer Reynolds numbers are  
defined as: 
\begin{flalign}
 \Pran&=\frac{\nu}{\kappa}, 
 \label{eq:Prandtl}\\
 \Rey&=\frac{U_\infty\delta}{\nu}=\sqrt{2\Rey_L},
 \label{eq:Reynolds}
\end{flalign}
where $\kappa$ is the molecular diffusivity 
of buoyancy and $\nu$ is the kinematic viscosity of abyssal seawater, and 
where the Stokes' layer thickness is:
\begin{equation}
 \delta=\sqrt{2\nu/\omega}.
 \label{eq:deltaS}
\end{equation} 
The slope frequency ratio is defined as 
the ratio of the projection of the buoyant acceleration onto the across slope ($x$) 
direction (parallel to the forcing) to the acceleration of the oscillatory forcing,
\begin{equation}
 \textit{C}=\frac{N\sin\theta}{\omega}.
 \label{eq:frequency_ratio}
\end{equation}
The slope frequency ratio was first identified as an important ratio for 
describing the boundary layer by \citet{Hart71} 
(who denoted it as $\mathrm{Q}$, where $\mathrm{Q}=\textit{C}^2$), 
while the frequency ratio 
$N/\omega$ emerges as an important measure of the role of 
stratification in the stratified form of Stokes' second problem 
(\citet{Gayen10b}). The slope frequency ratio 
is indicative of the degree of resonance between the oscillation 
body forcing  
and the buoyant restoring force.

Finally, the fourth non-dimensional ratio is 
slope Rossby number,
\begin{equation}
 \mathrm{Ro}=\frac{\omega}{f\cos\theta},
 \label{eq:Rossby}
\end{equation}
which indicates the ratio of the influence of 
planetary vorticity (projected onto the wall normal direction)
relative to vorticity with a characteristic time scale of 
the tide period, $\omega^{-1}$. For the finite Rossby number 
cases examined,
$\omega$ is the M$_2$ tide frequency, the
Coriolis parameter, $f$, is $10^{-4}$ s$^{-1}$, and 
the range of 
slope angles investigated are within $0<\theta\leq14^\circ$.
Therefore, the slope Rossby 
number is approximately 1.4 for all of the rotating 
reference frame flows investigated. 

\subsection{Invisicid, linear flow}
The inviscid, linearized form of the governing equations 
(Equations \ref{eq:continuity}-\ref{eq:buoyancy_nondim}) 
describe the
heaving of isopycnals up and down the slope,
\begin{flalign}
  \partial_t\tilde{{u}}
  &=\frac{1}{\textit{Ro}}\tilde{v}+\textit{C}^2\tilde{b}
  +F(t),
  \label{eq:inviscid_x}\\
   \partial_t\tilde{{v}}
  &=-\frac{1}{\textit{Ro}}\tilde{u},\\
  \partial_t\tilde{{b}}
 &=-\tilde{{u}}.\label{eq:inviscid_b}
\end{flalign}
Crucially, the solutions to Equations \ref{eq:inviscid_x}-\ref{eq:inviscid_b}
\textit{prescribe} the amplitude of the non-dimensional body force
\begin{flalign}
    A&=\Big( 
    \textit{C}^2+\frac{1}{\textit{Ro}^{2}}-1\Big),\label{eq:Asoln} 
\end{flalign}
where $A_\mathrm{d}=AU_\infty\omega$. 
The solutions to 
Equations \ref{eq:inviscid_b}-\ref{eq:inviscid_b} are
\begin{flalign}
  \tilde{{u}}(t)&=-\cos(t), \label{eq:zinf_ubc} \\
  \tilde{{v}}(t)&=\textit{Ro}^{-1}\sin(t), \label{eq:zinf_vbc}\\
  \tilde{{b}}(t)&=\sin(t).
\end{flalign}

\subsection{Resonance}
The amplitude of the oscillating 
pressure gradient forcing $F(t)$ in equation \ref{eq:inviscid_x} 
(the last term on the right hand side), 
vanishes if the critical 
slope condition
\begin{equation}
 \textit{C}^2+\frac{1}{\textit{Ro}^{2}}-1=0,
 \label{eq:criticality}
\end{equation}
is satisfied. 
Furthermore, if 
Equation \ref{eq:criticality} is satisfied, 
the energy of the inviscid baroclinic tide is tightly focused 
into narrow beams that follow 
the curvature of the bathymetry (\citet{Balmforth02});
therefore, the assumption of locally irrotational flow breaks down 
at critical slope, even 
on length scales much less than $L$.
At critical slope 
Equation \ref{eq:forcing} is no longer a valid approximation 
of the low-wavenumber baroclinic response to the barotropic tidal forcing.

Equation \ref{eq:criticality} is formally consistent with 
internal wave theory.
In internal wave theory, critical slopes are defined by: 
\begin{equation}
 \tan\theta_c=\sqrt{\frac{\omega^2-f^2}{N^2-\omega^2}},
 \label{eq:theta_c}
\end{equation}
where $\theta_c$ is the critical slope angle. 
If the slope angle $\theta\neq\theta_c$, then
Equation \ref{eq:Asoln} is satisfied, $A\neq0$. Equation 
\ref{eq:Asoln} can be rearranged to obtain
\begin{equation}
 \tan\theta=\sqrt{\frac{\omega^2(1+A)-f^2}{N^2-\omega^2(1+A)}}.
 \label{eq:theta}
\end{equation}
Therefore the criticality condition in Equation \ref{eq:criticality} 
is just a rearrangement of the criticality condition defined by 
the slope parameter $\epsilon$ from internal wave theory:
\begin{equation}
 \epsilon=\frac{\tan\theta}{\tan\theta_c},
 \label{eq:slope_parameter}
\end{equation}
where criticality states are defined:
\begin{flalign}
 \text{if}&\quad A<0\quad\text{then}\quad\epsilon<1
 \qquad\rightarrow\qquad\theta\hspace{1.7mm}\text{is subcritical},\nonumber\\
 \text{if}&\quad A=0\quad\text{then}\quad\epsilon=1
 \qquad\rightarrow\qquad\theta\hspace{1.7mm}\text{is critical},\nonumber\\
  \text{if}&\quad A>0\quad\text{then}\quad\epsilon>1
 \qquad\rightarrow\qquad\theta\hspace{1.7mm}\text{is supercritical}.\nonumber
\end{flalign}

\subsection{Boundary conditions}
At the solid boundary at $z=0$,
the boundary conditions on the total velocity are no-slip and 
impermeability
\begin{equation}
 \tilde{\mathbf{u}}=0,
 \label{eq:noslip}
\end{equation}
and the boundary conditions on the total buoyancy is 
the adiabatic condition:
\begin{equation}
 \partial_z\tilde{b}=0. 
 \label{eq:adiabatic}
\end{equation}
As $z\rightarrow\infty$, 
the velocity boundary conditions are the oscillatory 
solutions for the inviscid flow, 
and zero flow in the 
wall normal direction:
Equation \ref{eq:zinf_ubc},
Equation \ref{eq:zinf_vbc}, and $\tilde{{w}}=0$.
The non-dimensional buoyancy field as $z\rightarrow\infty$ has two 
components, the inviscid oscillation and the 
constant background stratification:
\begin{equation}
 \tilde{{b}}=x+z\cot\theta+\sin(t). 
 \label{eq:zinf_b}
\end{equation}

\subsection{Variable decomposition}
The
total velocity and buoyancy fields 
are decomposed into three components that when summed together 
satisfy Equations \ref{eq:continuity}-\ref{eq:buoyancy_nondim} 
and \ref{eq:noslip}-\ref{eq:zinf_b}.
To distinguish the components, let ``H'' denote the 
hydrostatic (and possibly geostrophic) 
component 1, let ``S'' denote the steady component 2, and let
``O'' denote the oscillating component:
\begin{flalign}
  \tilde{\mathbf{u}}(x,y,z,t)&=
  \mathbf{u}_{\text{H}}+\mathbf{u}_{\text{S}}(z)+\mathbf{u}_{\text{O}}(x,y,z,t),
  \label{eq:u_decomp}
  \\
  \tilde{b}(x,y,z,t)&=b_{\text{H}}(x,z)+b_{\text{S}}(z)+b_{\text{O}}(x,y,z,t),
  \label{eq:b_decomp}\\
  \tilde{p}(x,y,z,t)&=p_{\text{H}}(x,z)+p_{\text{S}}(z)+p_{\text{O}}(x,y,z,t).
  \label{eq:p_decomp}
\end{flalign}
The hydrostatic component of the buoyancy field  is merely the background 
stratification in the rotated coordinate system,
$b_{\text{H,d}}(x_\mathrm{d},z_\mathrm{d})
=N^2(x_\mathrm{d}\sin\theta+z_\mathrm{d}\cos\theta)$ in dimensional form, 
and the hydrostatic velocity field is zero everywhere except for the  
finite Rossby number flow regime, 
in which case it is the along-slope 
geostrophic velocity that arises from 
the across-isobath pressure gradient (\citet{Phillips70}, \citet{Wunsch70}). The buoyancy frequency $N$
is defined in the same manner as convention,
\begin{equation}
 N^2=-\frac{g}{\rho_0}\partial_\eta{\rho}_{\text{H}}=\partial_\eta{b}_{\text{H}}, 
 \label{eq:buoyancy_frequency}
\end{equation}
where $\eta$ denotes the vertical position coordinate. 

The steady and oscillating flow components are 
anomalies 
from the hydrostatic background 
that ensure the satisfaction of frictional and diffusive boundary conditions at the wall 
and inviscid oscillations far from the wall. 
It is convenient to 
solve for the anomalies,
\begin{flalign}
 \mathbf{u}(x,y,z,t)&=
 \mathbf{u}_{\text{S}}(z)+\mathbf{u}_{\text{O}}(x,y,z,t),
 \label{eq:u_anomaly_def}\\
  b(x,y,z,t)&=b_{\text{S}}(z)+b_{\text{O}}(x,y,z,t),
  \label{eq:b_anomaly_def}\\
  p(x,y,z,t)&=p_{\text{S}}(z)+p_{\text{O}}(x,y,z,t).
  \label{eq:p_anomaly_def}
\end{flalign}
because the removal of the hydrostatic background 
permits periodic analytical and numerical solutions 
for $\mathbf{u}$ and $b$. 

\section{Linear solutions}
Analytical
solutions to linearized forms of Equations \ref{eq:continuity}-\ref{eq:buoyancy_nondim}
contain a wealth of information pertaining to 
the laminar, disturbed laminar, 
and intermittently turbulent regimes 
(i.e. low to moderate Reynolds number flows)  
that are investigated numerically in this study.
\citet{Thorpe87} provided solutions of the 
rotating linear problem, a detailed derivation of which is given in Appendix~A. The solutions in Appendix~A are written in a form that readily collapses in 
the $\textit{Ro}\rightarrow\infty$ regime.

\subsection{Necessary conditions for gravitational instability}
The linear solutions for the steady flow component of both rotating and non-rotating flows 
is always gravitationally stable, meaning that 
the total vertical buoyancy gradient 
is never negative,
\begin{equation}
 \partial_{\eta}b_\mathrm{H}+\partial_{\eta}b_\mathrm{S}
 =1-\cos^2\theta\mathrm{e}^{-\eta/\delta_\mathrm{S}}
 \sqrt{2}\sin\Big(\frac{\eta}{\delta_\mathrm{S}}+\frac{\pi}{4}\Big)\geq0,
 \label{eq:static_stable}
\end{equation}
if the oscillating component vanishes. 
The 
boundary layer thickness of the steady component
\citet{Phillips70,Wunsch70} is
\begin{equation}
 \delta_\mathrm{S}=\Big(\frac{f^2\cos^2\theta}{4\nu^2}
    +\Pran \frac{N^2\sin^2\theta}{4\nu^2}\Big)^{-1/4}.
\end{equation}
The oscillating flow component solutions 
exhibit transient gravitationally unstable buoyancy gradients, $\delta_\eta\tilde{b}<0$, 
when denser fluid is advected over lighter 
fluid during portions of the oscillation period.
If the oscillating component is non-zero, then the minimum 
necessary condition for gravitational instabilities is 
defined by
\begin{equation}
  \partial_{\eta}b_\mathrm{H}+\partial_{\eta}b_\mathrm{S}<-\partial_{\eta}b_\mathrm{O},
  \label{eq:static_unstable}
\end{equation}
because if Equation \ref{eq:static_unstable} is 
satisfied, then the total vertical buoyancy gradient is negative,
$\delta_\eta\tilde{b}<0$.
However, instabilities can grow only if the 
negative buoyancy gradient is
sustained for a significant amount of time and 
if it is negative enough to
overcome 
resistance from friction.

A characteristic boundary layer 
Rayleigh number and a ratio of the time scale of the growth of an instability 
to the period of the oscillation are 
required to estimate the minimum (quasi-steady) 
conditions for the \textit{growth} of gravitational 
instabilities. However, the linear solutions 
do not readily yield a 
single boundary layer buoyancy gradient length scale. 
If the buoyancy gradient length scale is assumed to 
scale with $\delta=\sqrt{2\nu/\omega}$, then a tenable 
time-dependent 
boundary layer 
Rayleigh 
number is defined
\begin{flalign}
 \textit{Ra}(t)&\sim
 \frac{4\Pran N^2}{\omega^2}\partial_\eta\tilde{b}(t),
 \label{eq:Rayleigh}
\end{flalign} 
which only applies when $\partial_\eta\tilde{b}(t)<0$.
To estimate the graviationally stability of the flow 
without explicitly accounting for the time dependence of the basic state 
(the quasi-steady assumption), the 
basic state of the flow cannot change more rapidly than
the growth rate of a gravitational instability. 
If the instabilities are ``slowly modulated'' by the basic state \citet{Davis76}),
the quasi-steady assumption is reasonable for stability analysis. 
The dimensional 
instantaneous growth rate of a gravitational instability can be estimated as 
\begin{equation}
 \sigma\sim\Imag\Big[\sqrt{\partial_\eta \tilde{b}N^2}\Big].
\end{equation}
If $|{\omega}/{\sigma}|\ll1$,
then the modulation by the basic state is sufficiently slow
for the growth of gravitational instabilities. 

\begin{figure}
  \centering
 \includegraphics[width=1\textwidth,angle=0]{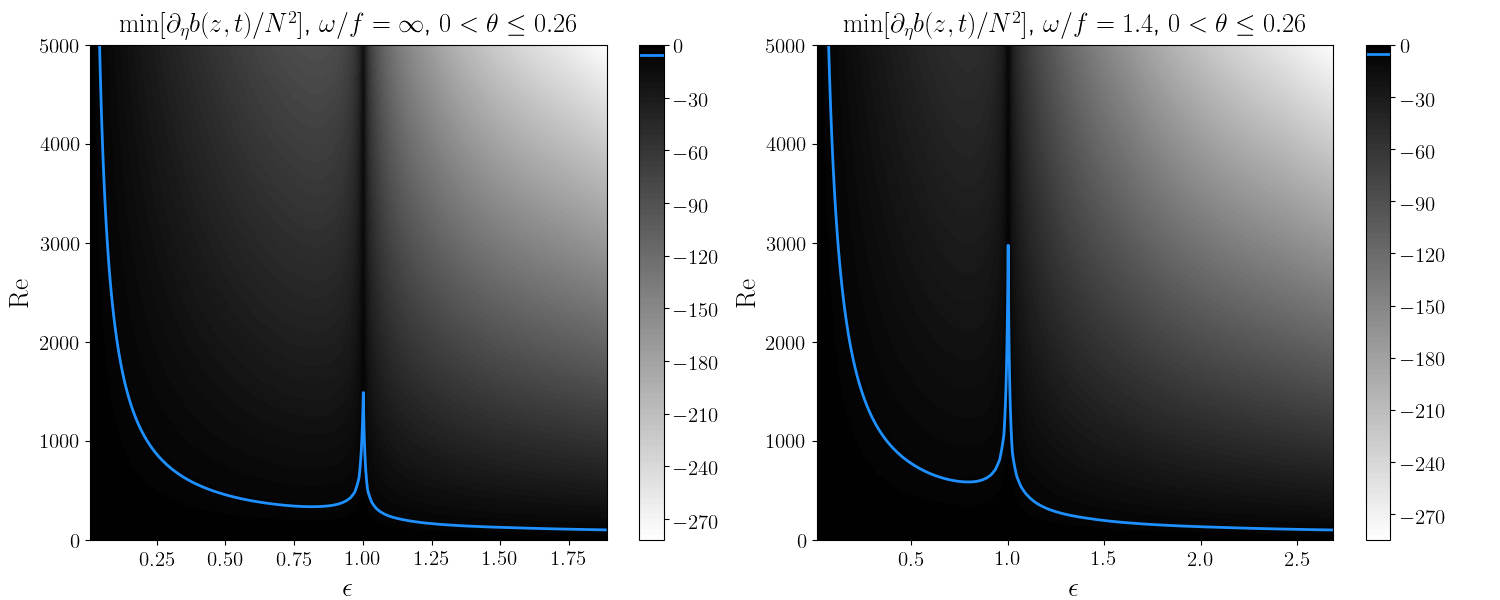}
 \caption{Total buoyancy gradient minima.} 
 \normalsize{\noindent The minimum value (in both time and space) of the 
 non-dimensional linear solution vertical buoyancy gradient 
 for the non-rotating reference frame case ($f=0$, \textit{left})
  
 and the rotating reference frame case (\textit{right}).
 }
 \label{fig:Rayleigh}
\end{figure}

Figure \ref{fig:Rayleigh} shows the minimum values of the 
non-dimensional total vertical buoyancy gradient 
from the linear solutions for non-rotating and rotating cases
as a function of slope parameter and 
Stokes Reynolds number.
Assuming that the gravitational instability in the boundary layer 
is physically similar to that of Rayleigh-B{\'{e}}nard instability in the case of 
one rigid and one stress-free boundary,
then the critical Rayleigh number  for the boundary layer 
is $\textit{Ra}_c\approx1100$ (\citet{Chandrasekhar61}),
which corresponds to $|\omega/\sigma|=0.06$. 
For the chosen fluid properties (holding $\Pran=1$, $f=10^{-4}$, $N=10^{-3}$, and 
$\omega=1.4\cdot10^{-4}$ constant), $\textit{Ra}_c\approx1100$ 
corresponds to a critical non-dimensional vertical 
buoyancy gradient of 
$\partial_\eta \tilde{b}=-5.4$, which is shown as the blue 
lines in Figure \ref{fig:Rayleigh}.
The minimum boundary layer buoyancy gradient is less than zero 
for all non-zero Re and $\epsilon$, and 
the minimum boundary layer buoyancy gradient is increasingly 
negative with increasing Reynolds number and 
with increasing slope parameter. The discontinuity at $\epsilon=1$ 
is an artifact of the degeneracy of linear solutions at critical slope.
The $\epsilon$ axis between the non-rotating and rotating cases is different 
because rotation alters the angle of critical slope; both 
plots show the same slope angle range, $0<\theta\leq16^\circ$. 

\section{Nonlinear solutions}
In this section we examine the nonlinear stability and 
development of turbulence in boundary layers on smooth 
abyssal slopes for both rotating and non-rotating regimes. The boundary layers are initialized by 
the oscillating laminar flow solutions derived in Appendix A.
We varied the slope Rossby number (nearly constant with slope, Equation \ref{eq:Rossby}), 
slope frequency ratio (Equation \ref{eq:frequency_ratio}),
Reynolds number (Equation \ref{eq:Reynolds}), 
slope parameter (Equation \ref{eq:slope_parameter}),
and slope angle $\theta$ for each of the 16 simulations 
as shown 
in table \ref{tab:nondimensional_parameters}. 
The slope frequency ratio $\textit{C}$ and slope parameter $\epsilon$ 
are redundant for the non-rotating case, 
but are shown together because $\textit{C}\neq\epsilon$ for the 
rotating flow, and $\textit{C}$ appears explicitly in the forcing 
of the across-isobath ($x$) momentum equation.
We observed bursts of turbulence, triggered by two-dimensional gravitational 
instabilities that rapidly become three-dimensional, during 
the upslope flow phase of all cases at $\Rey=840$ except 
for the lowest slope Burger number case, which exhibits turbulence sustained throughout the period. At $\Rey=420$ the 
flow matched the laminar analytical solutions except for weak 
turbulent bursts that occurred at the highest slope angles in the 
rotating regime.

\begin{table*}
\center{
\begin{tabular}{ l c c c c r }
\hline 
         &         &             &              &              &                         \vspace{-1.5mm} \\
 Cases   & Re      & Ro          & $\textit{C}$ &  $\epsilon$ & $\theta$ (rad)     \vspace{2mm} \\
 1,9     & 840,420 & $\infty$    & 0.25         &  0.25        & 3.53$\cdot10^{-2}$ \\
 2,10    & 840,420 & $\infty$    & 0.75         &  0.75        & 1.06$\cdot10^{-1}$ \\
 3,11    & 840,420 & $\infty$    & 1.25         &  1.25        & 1.76$\cdot10^{-1}$ \\
 4,12    & 840,420 & $\infty$    & 1.75         &  1.75        & 2.47$\cdot10^{-1}$ \\
 5,13    & 840,420 & 1.41       & 0.25         &  0.35        & 3.53$\cdot10^{-2}$ \\
 6,14    & 840,420 & 1.41       & 0.75         &  1.06        & 1.06$\cdot10^{-1}$ \\
 7,15    & 840,420 & 1.43       & 1.25         &  1.79        & 1.76$\cdot10^{-1}$ \\
 8,16    & 840,420 & 1.45       & 1.75         &  2.53        & 2.47$\cdot10^{-1}$ \\ 
         &         &             &              &              &           \vspace{-1.5mm} \\
  \hline 
\end{tabular}}
\caption{Non-dimensional simulation parameters.}
\normalsize{\noindent The four independent parameters are $\Rey$,
$\mathrm{Ro}$, $\textit{C}$, $\theta$, and $\Pran=1$. 
The slope parameter $\epsilon=\tan\theta/\tan\theta_c$ is also used in this 
study to directly connect results to internal wave parameters. The slope 
Burger number is $\mathrm{Bu}={N^2\tan^2\theta}/f^2=\mathrm{Ro}^2\textit{C}^2$.}
\label{tab:nondimensional_parameters}
\end{table*}

\subsection{Numerical implementation}
The flow anomalies, as defined by equations \ref{eq:u_anomaly_def} and \ref{eq:b_anomaly_def}, 
are discretized to
satisfy periodic boundary conditions in the wall parallel directions
via
Fourier spectral bases in the across-isobath ($x$) and along-isobath ($y$) 
directions. 
Periodicity is not merely numerically convienent; it 
also eliminates
the need to prescribe buoyancy forcing (``restratification'') 
because the oscillating flow can advect 
the background field to gain or lose buoyancy. 
In the periodic domain, 
the boundary layer buoyancy can only reach a homogenized steady 
state if the turbulence 
sustainably 
converts tidal momentum to potential energy throughout 
the entire period.

Although
the planar extent of the computational 
domain is less than the excursion length of the tide, 
the domain size (Table \ref{tab:grid}) 
is 
justifiably sufficient because 
the largest eddies in oscillating boundary layers 
are those associated with the transverse (wall normal) length scale,
which is much less than the excursion length.  
Indeed, at higher Reynolds number ($\Rey=1790$), \citet{Gayen10}
found the turbulent boundary layer thickness, $\delta_l$, was $\delta_l=15\delta$ for the unstratified problem
and $\delta_l=17\delta$ for flat plate stratified oscillating boundary layers at the same Reynolds number. 
The grid resolution parameters for the two Reynolds numbers examined 
are shown in Table \ref{tab:grid}, where ($L_x,L_y,H$) are the domain 
dimensions in ($x,y,z$),
$l_\text{K}$ and $\tau_\text{K}$ are the Kolmogorov length and time scales, respectively, 
and wall units (denoted by $^+$) are scaled by the viscous length scale $\delta_v=\nu/U_*$ where
$U_*$ is the a priori estimate of the friction velocity, which is approximated as 
$U_*=\sqrt{\nu\partial_z\overline{u}}\sim\sqrt{\nu{U_\infty}/{\delta}}$.

Solutions to the nonlinear anomaly equations, 
\begin{flalign}
  \partial_x{u}&+\partial_y{v}+\partial_z{w}=0,
  \label{eq:continuity_anomaly}\\
  \mathrm{d}_t{{u}}
  &=\frac{1}{\textit{Ro}}{v}-\partial_x{p}+
  \frac{1}{\Rey_L}\big(\partial_{xx}+\partial_{yy}+
  \partial_{zz}\big){u}+\textit{C}^2{b}
  +F(t)
  \label{eq:xmomentum_anomaly}\\
   \mathrm{d}_t{{v}}
  &=-\frac{1}{\textit{Ro}}{u}-\partial_y{p}+
  \frac{1}{\Rey_L}\big(\partial_{xx}+\partial_{yy}+\partial_{zz}\big){v},\\
  \mathrm{d}_t{{w}}
  &=-\partial_z{p}+
  \frac{1}{\Rey_L}\big(\partial_{xx}+\partial_{yy}+\partial_{zz}\big){w}
  +\textit{C}^2{b}\cot\theta,\\
  \mathrm{d}_t{{b}}
  &=
    \frac{1}{\Pran \hspace{0.3mm}\Rey_L}
    \big(\partial_{xx}+\partial_{yy}+\partial_{zz}\big){b},
    \label{eq:buoyancy_nondim_anomaly}
\end{flalign}
where the anomalies $\mathrm{u},p$ and $b$ 
are defined by Equations \ref{eq:u_anomaly_def}-\ref{eq:p_anomaly_def},
were computed using the MPI-parallel pseudo-spectral partial differential equation 
solver \textit{Dedalus} 
(\citet{Burns19}) using  $128^3$ modes. 
A third-order, four-stage, implicit-explicit 
Runge-Kutta method derived by \citet{Ascher97} was used for temporal integration.
Chebyshev polynomial bases of the first kind were employed for spatial 
discretization on a cosine grid in the
wall normal direction. 
Chebyshev polynomials permit the exact enforcement of the
adiabatic wall boundary condition 
(Equation \ref{eq:adiabatic} minus the background component) 
on the buoyancy field and no-slip/impermeability wall 
boundary conditions on the velocities 
(Equation \ref{eq:noslip} minus the background component). 
The $3/2$ rule dealiasing scheme is used not only for 
dealiasing the spatial modes online but also for dealiasing post-processed 
flow statistics.

At the maximum wall normal extent of 
the domain, 
the boundary conditions at infinity (Equations 
\ref{eq:zinf_ubc}, \ref{eq:zinf_vbc} and \ref{eq:zinf_b})  
were approximated 
for the anomalies 
as free-slip, impermeable conditions
\begin{equation}
  \partial_zu=\partial_zv=w=0,
\end{equation}
and an adiabatic condition on just the anomaly:
\begin{equation}
 \partial_zb=0,
\end{equation}
such that the total flow buoyancy gradient at the $z=H$ is the background 
buoyancy gradient in that direction. 

Although the impermeability 
condition causes the reflection of internal waves that reach the upper boundary,
the effects are assumed to be negligible because of the 
negligible amount of energy propagated by such high wavenumber 
waves in low Reynolds number flow. 
\citet{Gayen10b} 
found that for flat-bottomed stratified oscillatory flow at 
larger Reynolds number flow ($\Rey=1790$),
the vertical wave energy flux is less than 1\% of the boundary layer 
dissipation and production rates.
Indeed, small but non-zero   
dissipation rates of turbulent kinetic energy were found near
the upper boundary in some of the simulations, presumably from 
subharmonic parameteric instability or other wave-wave instabilities 
because of the free-slip reflective upper boundary condition. However, 
99.9\% of the shear production rate and dissipation rate 
occured within one Ozmidov length of the wall at the lower boundary for all simulations.

The initial conditions were specified as the sum of the 
steady component (\citet{Phillips70},\citet{Wunsch70}), 
the  
oscillating component (\citet{Thorpe87}) at time $t=0$,
and uniformly 
distributed white noise 
corresponding to buoyancy anomaly perturbations
of magnitude $10^{-10}$ m s$^{-2}$. 
All of the 
simulations that exhibited turbulence (defined as wall normal integrated production 
rates of turbulent kinetic energy greater than $10^{-10}$ m$^{3}$ s$^{-3}$) 
did so within two 
oscillations.

The parameter regimes show in Table \ref{tab:nondimensional_parameters} qualitatively describe 
flows forced by
the 
$M_2$ tide frequency, which is specified as
$\omega=1.4\cdot10^{-4}$ (rad$\hspace{0.5mm}$s$^{-1}$, for a 12.4 hr tide period) 
and the 
Coriolis parameter is specified as 
is $f=10^{-4}$ (rad$\hspace{0.5mm}$s$^{-1}$).
Much of the abyssal ocean is filled with 
Antarctic Bottom Water (AABW), characterized by 
temperatures near $0^\circ{}$C and 
practical salinities of 
approximately 35 psu. At $0^\circ{}C$ and 35 psu, the kinematic viscosity 
is $1.83\times10^{-6}$ m$^{2}$ s$^{-1}$ and the thermal diffusivity 
is $1.37\times10^{-6}$ m$^{2}$ s$^{-1}$ (\citet{Chen73}, \citet{Talley11}).
We specify the kinematic viscosity as  
$\nu=2\times10^{-6}$ m$^{2}$ s$^{-1}$ and Pr$=1$.
We approximate the background 
buoyancy frequency at mid latitude abyssal 
depths as $N=10^{-3}$ rad s$^{-1}$ (\citet{Thurnherr03}).
Baroclinic tide amplitudes 
of $U_{\scalebox{.9}{$\scriptscriptstyle \infty$}}\approx1$ cm s$^{-1}$ 
are routinely observed near abyssal slopes 
(\citet{Simmons04b}, \citet{Carter08}, \citet{Goff10}, \citet{Turnewitsch13})
far from critical slopes, hydraulic spills, and other turbulence ``hot spots.'' \citet{Becker08}  showed that the mean slopes
lie between 0.00 and 0.10 over the vast majority of the seafloor below 2000 m,
and that up to 15\% of the 
slopes are super critical with respect to the M$_2$ tide. 

\begin{table*}
\center
\begin{tabular}{ l c c c c c c c r }
\hline 
      &                           &            &                            &                           &                                       &               \vspace{-1.5mm} \\
 Re   & $L_x/\delta$,$L_y/\delta$ & $H/\delta$ &  $\Delta{}x^+,\Delta{}y^+$ & $\Delta{}z^+_\text{wall}$ & $\Delta{z}_\text{wall}/l_\text{K}$ & $\Delta{t}_\mathrm{d}/\tau_\text{K}$ & $\Delta{t}_\mathrm{d}/T$  \vspace{2mm} \\
 420  & 59.3                      & 177.8      &  9.5                       & 0.69                      & 0.40                                  & 0.01               & $2.2\cdot10^{-5}$  \\
 840  & 59.3                      & 177.8      &  13.5                      & 0.97                      & 0.52                                  & 0.02               & $2.2\cdot10^{-5}$         \\
      &                           &            &                            &                           &                                       &        \vspace{-1.5mm} \\
  \hline 
\end{tabular}
\caption{Simulation parameters}
\label{tab:grid}
\end{table*}

\subsection{Intermittent turbulent bursts}
The integrated turbulent kinetic energy (TKE) budget of each simulation was 
computed in order to distinguish the laminar and turbulent regimes
and to quantify turbulence production mechanisms.
The planar mean TKE is defined as
\begin{equation}
 K(z,t)\equiv\frac{1}{2}\big(\overline{u'^2}+\overline{v'^2}+\overline{w'^2}\big),
 \label{eq:tke_def}
\end{equation}
where the planar mean operator and variable decomposition are defined:
\begin{flalign}
\overline{\phi}(z,t)&=
 \frac{1}{L_xL_y}\int_{-L_x/2}^{L_x/2}\int_{-L_y/2}^{L_y/2}
 \phi(y,z,t)
 \hspace{1mm}\mathrm{dy}\hspace{1mm}\mathrm{dx},\\
 \phi(x,y,z,t)&=\overline{\phi}(z,t)+\phi'(x,y,z,t),
\end{flalign}
and $\phi$ is any of the anomalous variables defined by 
Equations \ref{eq:u_anomaly_def}-\ref{eq:p_anomaly_def}.
The planar mean TKE evolution equation is
\begin{equation}
  {\partial_t{K}}
  +\partial_{z}\mathcal{T}
 =\mathcal{P}
 +\mathcal{B} 
 -\varepsilon
 \label{eq:TKE_equation}
\end{equation}
The TKE transport term $\partial_{z}\mathcal{T}$ 
includes all TKE flux divergences (mean, turbulent, pressure, diffusion),
which vanish upon wall normal integration of equation 
\ref{eq:TKE_equation}. The rate production of TKE by mean shear is 
$\mathcal{P}$ 
(production in the sense that, generally, $\mathcal{P}>0$), and it is defined as
\begin{flalign}
 \mathcal{P}(z,t)&=
  -\overline{u'w'}\partial_z\overline{u}
  -\overline{v'w'}\partial_z\overline{v},\\
  \mathcal{P}_{13}&=-\overline{u'w'}\partial_z\overline{u},\\
  \mathcal{P}_{23}&=-\overline{v'w'}\partial_z\overline{v}.
\end{flalign}
The buoyancy flux $\mathcal{B}$ is typically downgradient 
($\mathcal{B}<0$ amidst $\partial_z\overline{b}>0$ or 
$\mathcal{B}>0$ amidst $\partial_z\overline{b}<0$), 
in which case it represents the conversion of 
TKE into potential energy, and it is defined as
\begin{flalign}
 \mathcal{B}(z,t)&=\overline{w_\eta'b'}
  =\overline{u'b'}\sin\theta+\overline{w'b'}\cos\theta,\\
 \mathcal{B}_1&=\overline{u'b'}\sin\theta,\\
 \mathcal{B}_3&=\overline{w'b'}\cos\theta,
\end{flalign}
in the rotated reference frame (where $w_\eta=\mathrm{d}_t\eta$ is the velocity 
in the vertical, not the wall normal velocity $w$).
Defined in this manner 
a downgradient buoyancy flux may be reversible, so the term 
$\mathcal{B}$ 
includes both the turbulent stirring of and turbulent diffusion of buoyancy.  
A reversible buoyancy flux may be thought of as a buoyancy flux that 
converts turbulent kinetic energy into potential energy through 
stirring alone.
Finally, the dissipation rate of turbulent kinetic energy, 
\begin{flalign}
  \varepsilon(z,t)=\hspace{1mm}&\nu\big(
   \overline{(\partial_x{u'})^2}
   +\overline{(\partial_x{v'})^2}
   +\overline{(\partial_x{w'})^2}
   +\overline{(\partial_y{u'})^2}
   +\overline{(\partial_y{v'})^2}\nonumber \\
   &+\overline{(\partial_y{w'})^2}
   +\overline{(\partial_z{u'})^2}
   +\overline{(\partial_z{v'})^2}
   +\overline{(\partial_z{w'})^2}\big)
   \label{eq:epsilon}
\end{flalign}
is positive definite and therefore the last term of equation \ref{eq:TKE_equation}
is always a sink of TKE.

The oscillating 
boundary layer buoyancy gradient is transient and evanescent; therefore, 
we seek an integral quantity to measure the 
stabilizing/destabilizing effects of the buoyancy gradient.
We borrow
the concept of boundary layer displacement thickness (\citet{Monin71})
and apply it to buoyancy gradients rather than momentum.
We refer to this measure as the boundary layer 
stratification thickness, $\delta_s$. 
If the total buoyancy field
is not constant over small distance in the wall normal direction $z_1$, 
then it can be approximated 
as constant over some distance $z_0$ from the wall, where $z_0$ 
is defined by
\begin{equation}
 N^2z_0=\int_0^{z_1}\partial_z\tilde{b}\hspace{1mm}\mathrm{dz}.
 \label{eq:Nintegral}
\end{equation}
It follows that
\begin{equation}
 \delta_s=z_1-z_0=\int_0^{z_1}\Big(1-\frac{\partial_z\tilde{b}}{N^2}\Big)\hspace{1mm}\mathrm{dz}.
\end{equation}
$z_1$ can be chosen arbitrarily as a point outside the boundary layer because 
outside of the boundary layer $\partial_z\tilde{b}=N^2$ (the only non-zero buoyancy gradient component outside the boundary layer is the background hydrostatic component). Therefore,
we set $z_1=H$ to calculate the stratification thickness.

\begin{figure}
 \centering
 \includegraphics[width=3in,angle=0]{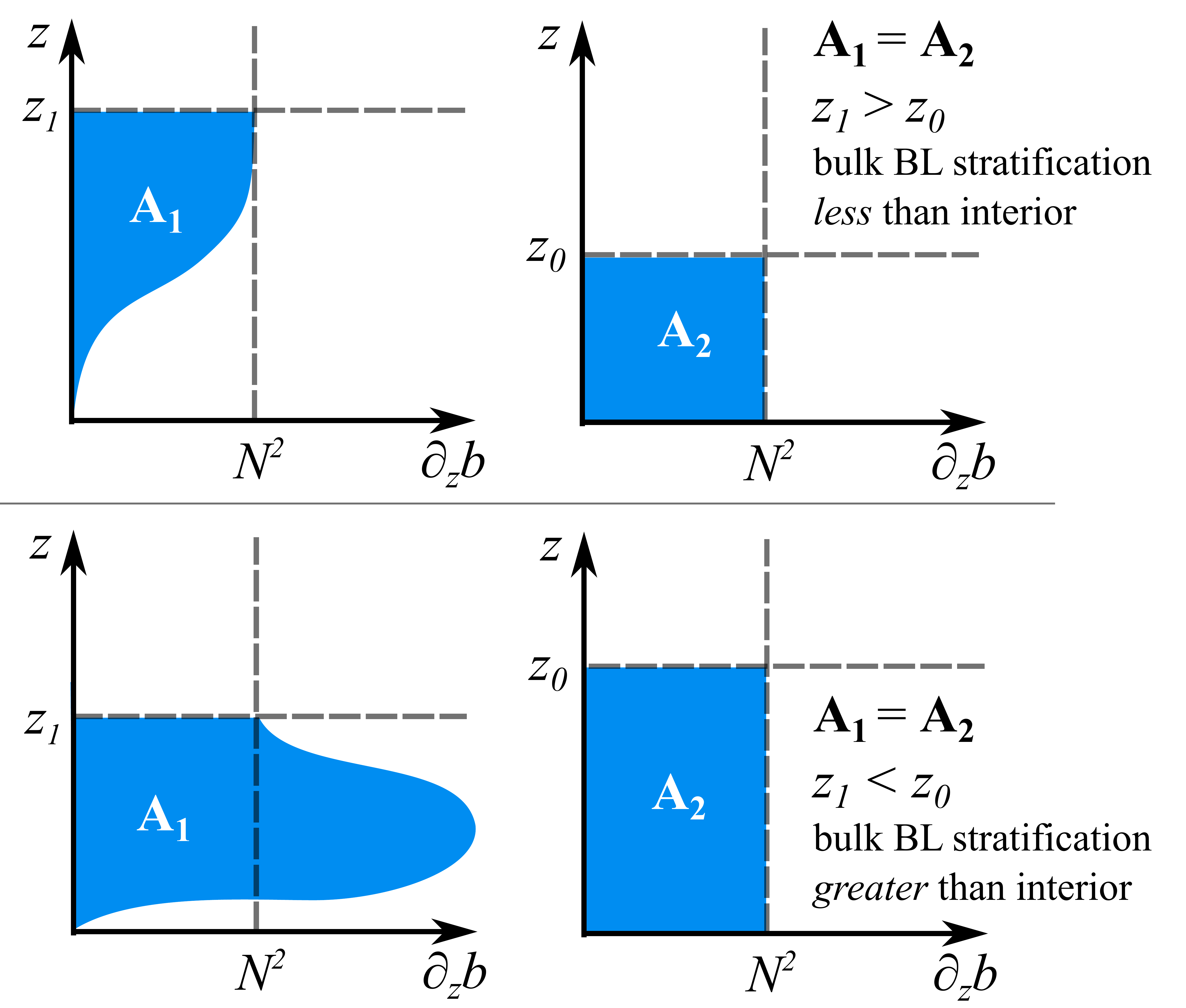}
 \caption{Stratification thickness concept. 
 }
 \label{fig:stratification_thickness}
\end{figure}

The physical interpretation of Equation \ref{eq:Nintegral} is 
illustrated in Figure \ref{fig:stratification_thickness}
as the calculation of the 
areas $\mathbf{A}_1$.
$\delta_s>0$ and indicates that, in 
bulk, the boundary layer stratification is less than the background stratification, 
and vice versa. 
$\delta_s<0$ for both 
the laminar steady flow components because the 
analytical solutions (\citet{Phillips70},\citet{Wunsch70}) prescribe positive bulk stratification near the boundary where isopycnals curve downwards.

\begin{figure}
 \centering
 \includegraphics[width=4.4in,angle=0]{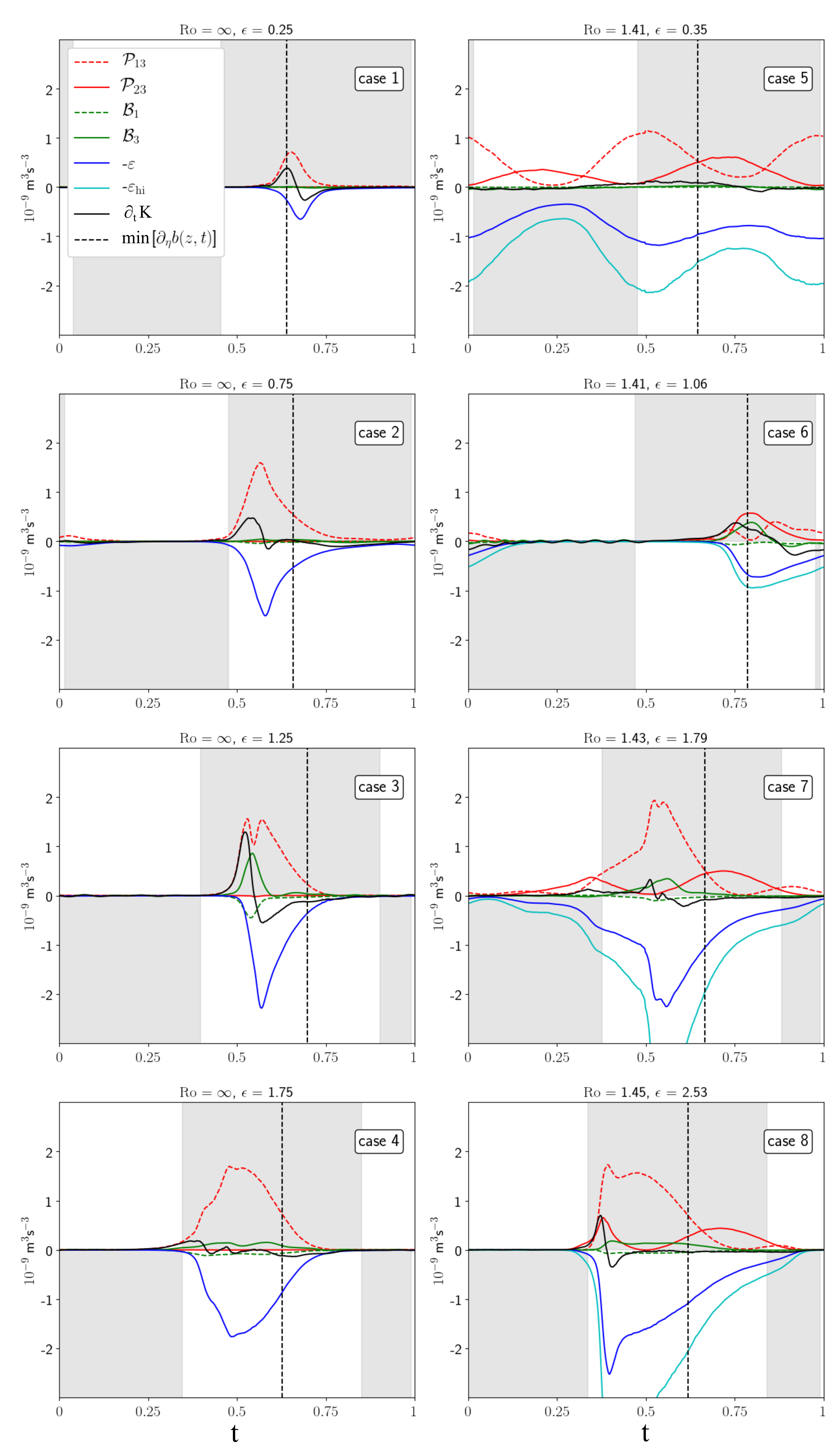}
 \caption{Wall-normal integrated,  planar mean
 TKE budgets.}
 \normalsize{\noindent The gray shading corresponds to the sign of the 
 stratification thickness (negative represents enhanced bulk boundary  
 layer stratification, postive represents weakened and/or negative bulk 
 boundary layer stratification. The dashed lines correspond to the time 
 of the minimum total vertical buoyancy gradient in the linear solutions.}
 \label{fig:fig_tke_Hov_6}
\end{figure}

The wave- and planar-averaged, wall normal integrated 
TKE budget statistics for $\Rey=840$
are shown in Figure \ref{fig:fig_tke_Hov_6}.  
The statistics were wave averaged over 5-10 oscillations.
All of the integrated TKE 
budgets at $\Rey=840$, 
with the exceptions 
of case 5 and arguably case 7, possess a single burst of chaotic three-dimensional 
motion that is
characterized by a rapid increase in the production rate of the TKE from the across-slope 
shear, $\mathcal{P}_{13}$, the component of shear parallel to the direction of the oscillating 
body force. 
The turbulent bursts, which occur shortly after $t/T\approx0.5$, 
preferentially select the phase regime during which the velocity is 
upslope but decelerating, the sign of the oscillating buoyancy  
changes from positive to negative, and the stratification thickness is negative 
(as indicated by the dark grey shading in Figure \ref{fig:fig_tke_Hov_6}). 
The negative stratification thickness preference of the bursts 
contrasts the low Reynolds number, intermittent turbulence 
regime of Stokes' second problem, in which a single burst 
occurs per oscillation, corresponding to the random selection of 
one of two 
shear maxima that occur within one period (\citet{Spalart87}). 

To investigate the role of the linear buoyancy dynamics in the formation of the 
turbulent bursts in Figure \ref{fig:fig_tke_Hov_6}, 
the time of the minimum total vertical buoyancy 
gradient in the linear solutions, which the reader may recall is negative for all of the 
considered parameter space as shown in figure \ref{fig:Rayleigh}, 
is depicted as the vertical dashed black 
line. 
The maximum TKE production rate by the mean shear 
approximately coincides in the time of the minimum total vertical 
buoyancy gradient for cases 1 and 6, the smallest intensity turbulent bursts
shown in Figure \ref{fig:fig_tke_Hov_6}.
This result
suggests that the bursts of TKE production rate by the mean shear 
are triggered by buoyant ejections of low momentum fluid upward. However, 
the triggering 
of buoyancy ejections is brief, weak, and not sustained, 
because the buoyancy fluxes in Figure \ref{fig:fig_tke_Hov_6} are negligible
prior to bursts in the along-isobath 
shear production. Thus the gravitational instabilities appear to 
initiate, but not drive, bursts of chaotic three-dimensional 
motion.

It is well known that boundary layer turbulence is inherently anisotropic. 
However, the majority of ocean turbulence measurements 
measure the fluctuations of the vertical shear of the horizontal velocities and then assume homogeneous isotropic turbulent motion to 
subsequently estimate the dissipation rate (\citet{Polzin96}, \citet{StLaurent01}). 
The
isotropic, homogeneous turbulence dissipation rate of TKE (\citet{Taylor35}) 
is defined in the sloped coordinate frame as
\begin{equation}
 \varepsilon_\text{hi}=
 \frac{15}{4}\nu\cos^2\theta\big(\overline{(\partial_zu')^2}
 +\overline{(\partial_zv')^2}\big).
 \label{eq:iso_dissip3}
\end{equation}
The wall normal integrated form of $\varepsilon_\mathrm{hi}$ 
are plotted for the rotating reference frame cases 
in Figure \ref{fig:fig_tke_Hov_6} to illustrate 
that the assumptions of homogeneous anisotropic turbulence 
for near-wall abyssal flows 
may lead to overpredictions of the dissipation rate of 
TKE on the order of 100\%, regardless of slope angle, even for the
low Reynolds number boundary layers investigated here.

\subsection{Gravitationally unstable rolls}
\begin{figure}
 \centering
 \includegraphics[width=5.25in,angle=0]{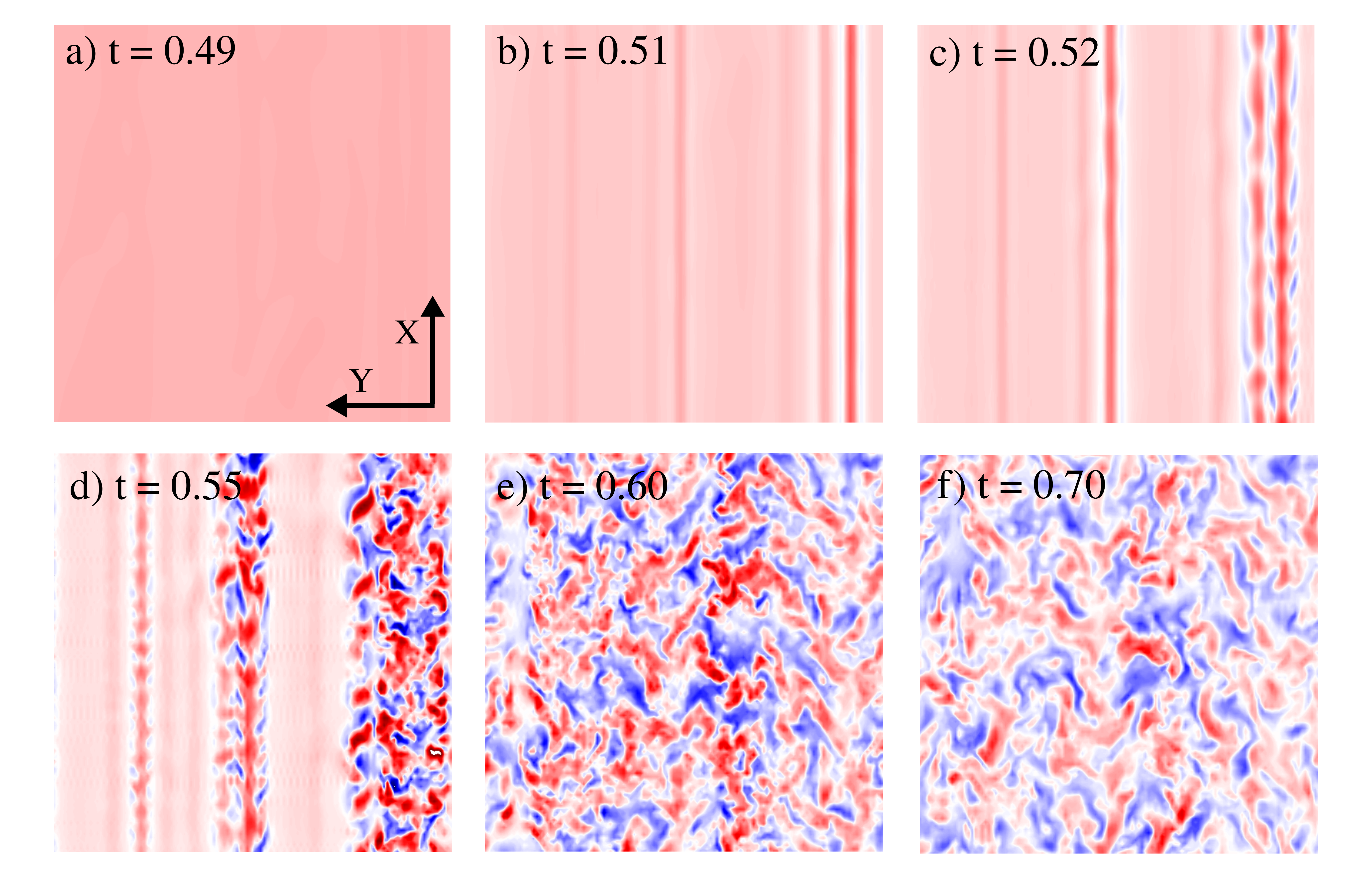}
 \caption{Contours of the vertical velocity, $w$, for case 2.}
 \normalsize{\noindent The contour plots show the 
 vertical velocity at a fixed distance (roughly $\delta$) 
 in the wall normal 
 direction at six consecutive times. $w>0$ is colored red, while $w<0$ is 
 colored blue. At $t=0.5$ the across-isobath velocity 
 is positive but begins to decelerate. Simultaneously, 
 two dimensional rolls form (see plot $b)$) in the $y-z$ plane,
 as heavier fluid is advected over lighter fluid trapped near the wall by 
 the friction.}
 \label{fig:ingersol}
\end{figure}

Except for case 6, all of the simulations at $\Rey=840$
exhibited rolls characterized by growing streamwise vorticity.
Figure \ref{fig:ingersol} 
shows the instantaneous vertical velocity 
of case 2 to illustrate the life cycle of the rolls. 
In Figure \ref{fig:ingersol}, red corresponds to upward motions, and 
blue approximately corresponds to downward motions. 
The generation of two-dimensional convective 
rolls in the along-isobath / wall-normal ($y$-$z$) plane 
are visible just prior to the beginning of a burst. At 
time $t=0.51$ the rolls appear and by $t=0.55$ the 
rolls have begun to shear apart, erupting 
into the three-dimensional turbulence at the time of
increase in TKE production by mean shear at $t=0.55$.

The rolls formed by gravitational instabilities 
in Figure \ref{fig:ingersol} are qualitatively consistent 
with rolls observed in 
oscillating sloping stratified boundary layer experiments by \citet{Hart71}.
\citet{Hart71} identified spanwise
plumes and rolls 
associated 
with the periodic reversals of the density gradient
that 
qualitatively resembled the rolls that appeared
in high Rayleigh number Couette flow experiments by 
\citet{Benard38}, \citet{Chandra38}, and \citet{Brunt51}.
Perhaps due to the similarity to the convection experiments,
the rolls observed by \citet{Hart71} were referred to
as ``convective rolls.'' 
Linear stability analyses by \citet{Deardorff65}, \citet{Gallagher65}, 
and \citet{Ingersoll66},
revealed that the growth of  
gravitationally unstable disturbances in
high Rayleigh number Couette flows
is suppressed in the plane of the shear (the streamwise-vertical plane) 
by the shear 
(i.e. the suppression of the spanwise vorticity 
disturbances). 
They also found that the growth of disturbances in the spanwise-vertical 
plane (steamwise vorticity 
disturbances) is unimpeded by the shear and grows in the same manner 
as pure convection. 
It has since been established that 
streamwise (the across-isobath direction)
vortices with axes in the 
direction of a mean shear flow (a.k.a. ``rolls'')
can arise due to heating or centrifugal effects (\citet{Hu97}).
The initial growth of the 
rolls in Figure \ref{fig:ingersol} appears to have similar attributes. 

To verify the hypothesis that gravitational instabilities 
spawn the rolls, which in turn spawn the turbulent burst, 
an additional simulation with the same parameters as that of
case 2, but with no nonlinear terms in the 
buoyancy equation, was executed. The simulation of the 
linearized buoyancy equation version 
of case 2 
had no turbulent bursts over
10 cycles (all other simulations with bursts developed a 
burst within 2 cycles).
The rolls are a bypass transition mechanism, lifting 
low momentum fluid up and bringing high momentum fluid down 
into the near wall flow, and so they transiently
destabilize the shear.
The transient gravitationally unstable buoyancy gradients, discussed 
previously, can 
trigger oscillating boundary layer turbulent bursts even if the buoyancy fluxes are  
a negligible source of TKE. 

Although the streamwise rolls are initially two dimensional, 
they produce a three dimensional vorticity field.
The inherent three dimensionality of the gravitational instability is evident
in the Boussinesq 
baroclinic production of vorticity term ($\nabla\times{}\tilde{b}$) in 
the absolute vorticity budget for rotating and non-rotating 
oscillating boundary layers, 
\begin{equation}
 \text{baroclinic production}
 =\textit{C}^2\big(\hspace{-4mm}\overbrace{\partial_y\tilde{b}\cot\theta\mathbf{i}}^{\substack{\text{nonlinear 2D rolls}\\\vspace{0.5mm}\text{\& 3D bursts}}}\hspace{-1.5mm}
 +\hspace{2.5mm}(\overbrace{\partial_x\tilde{b}\cot\theta}^{\substack{\text{nonlinear}\\\vspace{0.5mm}\text{3D bursts}}}\hspace{2mm}-\hspace{-2.5mm}\overbrace{\partial_z\tilde{b}}^{\substack{\text{linear flow}\\\vspace{0.5mm}\text{\& 3D bursts}}}\hspace{-4mm})\mathbf{j}\hspace{4mm}
 -\hspace{-4mm}\overbrace{\partial_y\tilde{b}\mathbf{k}}^{\substack{\text{nonlinear 2D rolls}\\\vspace{0.5mm}\text{\& 3D bursts}}}\hspace{-7.5mm}\big).
 \label{eq:baroclinic_production}
 \end{equation} 
The linear OBL vorticity field has only one vorticity component, the 
spanwise vorticity in the $y$ direction, and the linear ROBL vorticity 
field is comprised of the spanwise vorticity and the streamwise vorticity 
in the $x$ direction. In either case, 
only the $\partial_z\tilde{b}\mathbf{j}$ term in Equation 
\ref{eq:baroclinic_production} is non-zero. 
However, the rolls produce gradients in 
the buoyancy field in the $y$ direction. 
The first and last terms on the righthand side of 
equation \ref{eq:baroclinic_production} indicate that the 
rolls in the $y-z$ plane will inevitably generate vorticity in 
the streamwise and wall normal directions;
therefore, the rolls 
must induce three-dimensional motion in the oscillating boundary layers.
Therefore 
the coherent structures shown in figure \ref{fig:ingersol},
which facilitate 
the transition to turbulence, must initiate secondary 
instabilities through three-dimensional baroclinic production of  
vorticity, a phenomenon that is widely observed 
in other stratified shear flow instabilities, 
e.g. \citet{Peltier03}.

\subsection{Relaminarization}
During the phase of enhanced boundary layer stratification relative to the 
background, 
$\delta_s<0$ (the white regions in Figure \ref{fig:fig_tke_Hov_6}), 
the boundary shear instabilities must overcome 
not only the stabilizing effect of increased stratification 
but also the stabilizing effect of the wall that is present 
regardless of phase. 
Linear stability 
analysis by \citet{Schlichting35} yielded a critical gradient 
Richardson number of 1/24 for a stratified 
Blasius boundary layer, notably lower than the Miles--Howard theorem threshold
for inviscid stratified shear. This suggests that the flow is 
stabilized with respect to shear perturbations during $\delta_s<0$.
However, if the flow is turbulent during the phase of $\delta_s<0$, 
the turbulence closure  
simulations for high Reynolds number by 
\citet{Umlauf11} suggest mixing is much more 
efficient, presumably because the turbulence intensity 
required to overcome 
the strengthened boundary layer stratification must be considerable.

Three other 
mechanisms contribute to the relaminarization of 
the turbulent bursts in Figure \ref{fig:fig_tke_Hov_6}.
First, the 
turbulent burst diffuses the mean shear and thus its primary 
energy source. Second,  
the tidal acceleration opposes 
the mean shear during the second half of the phase, 
so the decay and reversal of 
the shear amplitude means that less mean flow kinetic energy is available.
Third, once the flow reverses the outer boundary layer 
becomes increasingly stratified, as mentioned previously, when $\delta_s<0$. 
For a burst to persist across the entire period it must have a 
constant source of mean shear of large enough magnitude to 
sustain production of TKE throughout flow reversals and 
increased stratification.

\begin{figure}
 \centering
 \includegraphics[width=5.8in,angle=0]{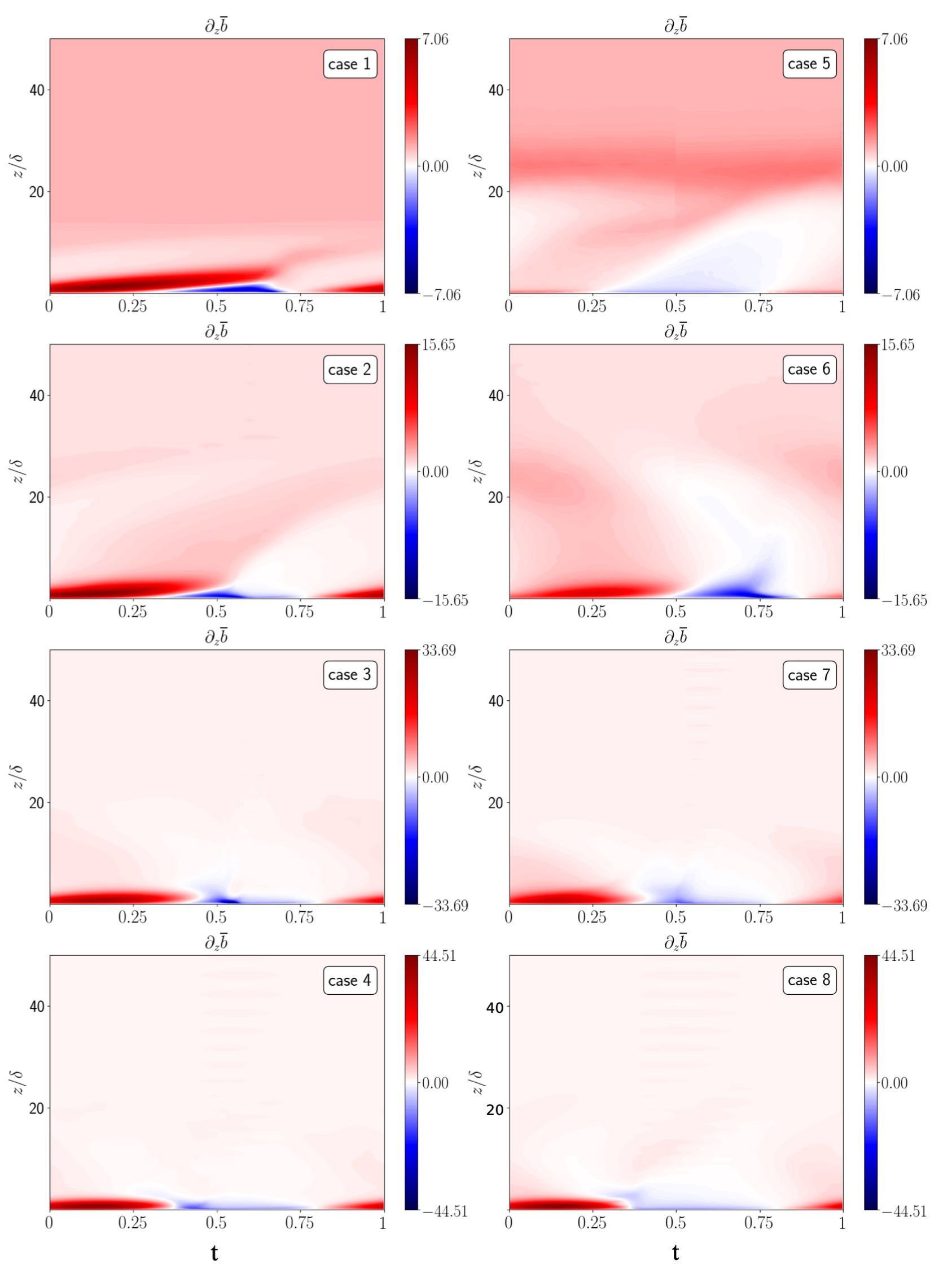} 
 \caption{
 $\Rey=840$ H{\"o}vmuller plots of mean stratification.}
 \normalsize{\noindent 
 The total wall normal buoyancy 
 gradients are non-dimensionalized by $N^2$. The color bar 
 axes show that the boundary layer stratification maxima/minima 
 increase/decrease with increasing slope.}
 \label{fig:tbz}
\end{figure}

\begin{table*}
\center
\begin{tabular}{ l c   }
\hline 
      &                                                         \vspace{-1.5mm} \\
 Case &   Bu   \vspace{2mm} \\
 5,13 &   0.124                                \\
 6,14 &   1.127                                 \\
 7,15 &   3.173                               \\
 8,16 &   6.347                             \\ 
      &                                             \vspace{-1.5mm} \\
  \hline 
\end{tabular}
\caption{Rotating boundary layer simulation slope Burger numbers.}
\label{tab:shutdown}
\end{table*}

\subsection{The effect of rotation}
Only case 5 features sustained turbulence throughout the period (Figure \ref{fig:fig_tke_Hov_6}) and sustained 
boundary layer buoyancy homogenization (Figure \ref{fig:tbz}), 
because projection of the Coriolis force onto the wall normal direction 
increases with decreases slope. 
The Burger number is the ratio of the squares of the inertial period to 
the time scale associated with the buoyant restoring force. 
The slope Burger number accounts for the rotated reference frame, defined as
\begin{flalign}
  \textit{Bu}&=\textit{Ro}^2\textit{C}^2=\frac{N^2\tan^2\theta}{f^2}.
 \label{eq:Burger}
\end{flalign}
If Bu $<1$, the buoyant restoring force acts more slowly than the Coriolis 
force and rotation is significant. 
Table \ref{tab:shutdown} shows the Burger number 
for 
the rotating boundary layer flows, cases 5--8 and 13--16. 
The lowest slope Burger number cases, 5 and 13, 
are influenced the most by rotation. In case~5, the highest 
Reynolds number and lowest Burger number flow, the turbulence is 
sustained throughout the period and the mean velocity field 
oscillates in a Stokes--Ekman layer manner. 
These results suggest that, within the investigated parameter 
regime, low Burger number flows are more likely to sustain 
turbulence throughout the entire period.

\begin{table*}
\center
\begin{tabular}{ l c  c }
\hline 
      &      &         \vspace{-1.5mm} \\
 Case & Re   & $c_D$   \vspace{2mm} \\
 1,9    & 840,420  & $1.8\cdot10^{-4}$, $8.8\cdot10^{-5}$    \\
 2,10    & 840,420  & $9.7\cdot10^{-4}$, $\sim0$                  \\
 3,11    & 840,420  & $1.6\cdot10^{-3}$, $\sim0$               \\
 4,12   & 840,420  & $3.5\cdot10^{-3}$, $\sim0$                \\ 
 5,13    & 840,420  & $4.7\cdot10^{-3}$, $\sim0$                                \\
 6,14    & 840,420  & $5.1\cdot10^{-4}$, $1.6\cdot10^{-8}$                                  \\
 7,15    & 840,420  & $5.0\cdot10^{-3}$, $1.5\cdot10^{-3}$                               \\
 8,16    & 840,420  & $5.4\cdot10^{-3}$, $1.4\cdot10^{-3}$                             \\ 
      &                                             \vspace{-1.5mm} \\
  \hline 
\end{tabular}
\caption{Drag coefficients estimated from simulation time mean dissipation rates.}
\label{tab:drag}
\end{table*}

\subsection{Barotropic tide dissipation}
In oceanography, the rate of energy loss to drag per m$^2$ of the barotropic flow by tidal bottom boundary layers 
is often estimated using the quasi-empirical model
$D\approx \rho_0c_D|U|U^2$ (\citet{Hoerner65}), where $c_D$ is the dimensionless drag coefficient and 
$U$ is an estimate of the bulk velocity
(\citet{Jayne01}, \citet{StLaurent02}). The rate 
of energy dissipated by the tide can be represented 
in terms of Watts per meter squared
by 
\begin{equation}
    D=\rho_0\int_0^H\overline{\varepsilon}\hspace{1mm}\mathrm{dz}, 
\end{equation}
where $\overline{\varepsilon}$ is the dimensional time mean dissipation rate of TKE (units m$^{2}$s$^{-3}$). 
For flat plate boundary layers, 
the transitional flow regime is characterized by drag coefficients 
in the range $0.001\leq{}c_D\leq0.005$, 
for $1<\Rey<10^3$ or equivalently $1<\Rey_L<10^6$ 
(\cite{Hoerner65}). The drag coefficients for the low Reynolds number 
boundary layers in this study were calculated as
\begin{equation}
 c_D=U_\infty^{-3}\int_0^H\overline{\varepsilon}\hspace{1mm}\mathrm{dz}.
\end{equation}

The drag coefficients are shown in Table \ref{tab:drag}. The drag coefficient 
values 
shown in Table \ref{tab:drag} mostly fall within the expected range for flat plate 
boundary layers, with the exception of the steepest slope case at $\Rey=840$, case 8.  
The drag coefficients are small at $\Rey=420$ for all but the 
steepest slope angles in the rotating reference frame (cases 15 and 16). 
In Table \ref{tab:drag}, $\sim0$ represents negligible drag. 
The drag coefficients increase with slope at constant Reynolds number, 
and they effectively vanish somewhere in the range $420<\Rey<840$ on 
lower slopes where the flow is in the laminar or disturbed laminar regime.

\section{Conclusions}
We investigated transition pathways of 
low Reynolds number, 
oscillating, stratified, diffusive boundary layers on 
infinite slopes in both rotating and non-rotating reference frames. Our goal was to answer three questions regarding 
these flows within a non-dimensional parameter 
space qualitatively consistent with mid- and low-latitude 
$M_2$ tide boundary layers on smooth slopes:
\begin{enumerate}
    \item \textit{How unstable are these boundary layers on typical abyssal slopes?} \\ Our results suggest that the laminar boundary layers are destabilized by a two-dimensional gravitational instability
    which rapidly progresses into three-dimensional motion and subsequently bursts of turbulence. This occurred for all of the investigated parameter space, except for at low slope Burger number (Bu$\leq1/8$) and for low Reynolds number, low slope angle (Re$=$420, $\theta<0.1$) cases.
    The low slope Burger number, $\Rey\approx840$, case is qualitatively similar to stratified Stokes-Ekman layers, and it was the only simulation 
that exhibited statistically stationary boundary layer buoyancy homogenization, or steady ``mixing.'' The low slope angle, low Reynolds number cases remained laminar.
    
    Vertically integrated TKE budgets suggest that energy supply to the turbulent bursts was extracted from the mean shear. 
    Increasing the slope 
parameter $\epsilon$ resulted in increases in turbulent burst intensity, quantified by integrated TKE shear production, and 
also induced larger positive turbulent 
buoyancy fluxes. The positive turbulent buoyancy 
fluxes eroded negative buoyancy gradients and generated downgradient buoyancy diffusion. The turbulent bursts in rotating reference frames (slope $\mathrm{Ro}=1.4$) 
were qualitatively similar to the non-rotating reference frame bursts. 
    
    During the phase of downslope mean flow the bursts where observed to relaminarize. With the exception of the low slope Burger number case, we observed that the boundary 
layer stratification significantly 
controls the transitional and intermittent 
regimes within $420\lesssim\Rey\lesssim840$ by suppressing turbulence during 
the downslope flow phase and triggering TKE production by 
the mean shear during the upslope flow phase.
    \item \textit{What are the instability mechanisms?}\\
    Bulk estimates of the maximum boundary layer Rayleigh number (see Figure \ref{fig:Rayleigh})
from analytical solutions were consistent with gravitationally instabilities
that were observed in direct numerical simulations of varying
parameter
space. 
The gravitational 
instabilities for rolls qualitatively resemble the convective rolls of 
diabatic Couette flow, and the correlation between the timing of 
their formation and gravitationally unstable stratfication in 
the boundary layer suggest that the instabilities are initiated 
by the upward ejection of buoyant low momentum fluid near the wall. Our results,
the linear instability of diabatic Couette flow (\citet{Ingersoll66}), and Floquet linear instability 
of the laminar flow solutions (\citet{kaiser2020finescale}), all suggest that the investigated flows are susceptible to graviational linear instability. 
    \item \textit{How valid are back-of-the-envelope barotropic tide dissipation rates?}\\
    The dissipation rates of TKE and drag coefficients increased with increased slope parameter, 
more for the non-rotating cases than the rotating cases. The drag coefficients are negligible for 
$\Rey\approx420$ flows, particularly on low angle slopes.
The drag coefficients (Table \ref{tab:drag}) become 
quite small as the Reynolds number is decreased from 840 to
420, which suggests that a low tidal Reynolds number cutoff  of $\Rey\approx840$
is appropriate for barotropic tide 
drag parameterizations. 
The drag coefficients increased with slope angle although the 
steepest slopes in this study are not found in the ocean at large 
scales (scales equal to or greater than $k^{-1}$, the 
horizontal length scale of bathymetry).
\end{enumerate}

\section{Acknowledgements}
B.K. was supported by a N.S.F. 
Graduate Research Fellowship and 
the Massachusetts Institute of Technology - Woods Hole 
Oceanographic Institution Joint Program, and
by the National
Science Foundation (OCE-1657870).
The authors 
thank the Massachusetts Green Computing Center, Keaton Burns, 
Jesse Canfield,
Raffaele Ferrari, Karl Helfrich, Kurt Polzin, Xiaozhou Ruan, and Andreas Thurnherr. 
This document is approved for Los Alamos Unlimited Release, LA-UR-21-28222.

\appendix
\section{}\label{appA}
The following is a derivation of the solutions to 
Equations \ref{eq:linear_xmomentum}, 
\ref{eq:linear_ymomentum}, and \ref{eq:linear_buoyancy}.
In the other chapters of this thesis, partial derivatives 
are denoted by $\partial_{zz}$ for the second derivative in 
$z$, for example. In this appendix, Leibniz notation is 
used for derivatives.
Begin by assuming \textit{linear} oscillating solutions of the form 
\begin{equation}
 u_{\textrm{O},d}=\mathcal{U}(z)\mathrm{e}^{\text{i}\omega{t}}, \quad
 v_{\textrm{O},d}=\mathcal{V}(z)\mathrm{e}^{\text{i}\omega{t}}, \quad
 b_{\textrm{O},d}=\mathcal{B}(z)\text{ie}^{\text{i}\omega{t}},
 \label{eq:ansatz}
\end{equation}
where $d$ denotes the variables are dimensional and $\mathrm{O}$ 
denotes the oscillating components (Equations \ref{eq:u_decomp} and 
\ref{eq:p_decomp}).
It does not matter if we make the ansatz 
$\mathcal{V}(z)\mathrm{e}^{\text{i}\omega{t}}$ or 
$\mathcal{V}(z)\text{ie}^{\text{i}\omega{t}}$ 
(the latter is the correct final form) 
because the particular solution  
fixes the phase relationship of $u$ and $v$.
The oscillating components 
of the dimensional and linearized forms of Equations \ref{eq:continuity}-\ref{eq:buoyancy_nondim}, 
with no variation in the 
across-isobath ($x$) or along-isobath ($y$) directions, satisfy
\begin{flalign}
  \partial_t{{u}_{\textrm{O},d}}
  &=f{v}_{\textrm{O},d}+
  \nu\partial_{zz}{u}_{\textrm{O},d}+{b}_{\textrm{O},d}\sin\theta
  +F_d(t)
  \label{eq:linear_xmomentum}\\
   \partial_t{{v}_{\textrm{O},d}}
  &=-f{u}+
  \nu\partial_{zz}{v}_{\textrm{O},d},
  \label{eq:linear_ymomentum}\\
  \partial_t{{b}_{\textrm{O},d}}
  &=-{u}_{\textrm{O},d}N^2\sin\theta+
    \kappa
    \partial_{zz}{b}_{\textrm{O},d},
    \label{eq:linear_buoyancy}
\end{flalign}
where the wall normal momentum vanishes by conservation of mass, and 
the wall normal momentum equation again reduces to a diagnostic equation 
for the pressure field. 
Substitution of the ansatz (Equations \ref{eq:ansatz}) into 
the linearized governing equations \ref{eq:linear_xmomentum}, 
\ref{eq:linear_ymomentum}, and \ref{eq:linear_buoyancy}
yields 
\begin{equation}
    \Big(\text{i}\omega
    -\nu\frac{\partial^2}{\partial{z^2}}\Big)\mathcal{U}
    =
    \mathcal{V}f\cos\theta
    +
    {i}\mathcal{B}\sin\theta
    +Ai,
    \label{eq:xmomentum_dim_linear2}
\end{equation}
\begin{equation}
\Big(\text{i}\omega
    -\nu\frac{\partial^2}{\partial{z^2}}\Big)\mathcal{V}
    =-\mathcal{U}f\cos\theta,
    \label{eq:ymomentum_dim_linear2}
\end{equation}
\begin{equation}
\Big(\text{i}\omega
    -\kappa\frac{\partial^2}{\partial{z^2}}\Big)
    i\mathcal{B}
    =-\mathcal{U}N^2\sin\theta.
    \label{eq:buoyancy_dim_linear2}
\end{equation}
The equations above can be reduced 
to a single inhomogeneous linear partial 
differential equation for the wall-normal buoyancy structure 
$\mathcal{B}(z)$: 
\begin{flalign}
    &\Bigg[\Big(\text{i}\omega
    -\nu\frac{\partial^2}{\partial{z^2}}\Big)\Big(\text{i}\omega
    -\nu\frac{\partial^2}{\partial{z^2}}\Big)\Big(\text{i}\omega
    -\kappa\frac{\partial^2}{\partial{z^2}}\Big)\nonumber 
     + 
        N^2\sin^2\theta\Big(\text{i}\omega
    -\nu\frac{\partial^2}{\partial{z^2}}\Big)\\
   & -f^2\cos^2\theta
\Big(\text{i}\omega
    -\kappa\frac{\partial^2}{\partial{z^2}}\Big)
    \Bigg]
    i\mathcal{B}
    =A\omega{}N^2\sin\theta. 
    \label{eq:xmomentum_dim_linear9}
\end{flalign}
Equation \ref{eq:xmomentum_dim_linear9} has 
six characteristic roots for the complementary
(homogeneous)
component of the solution and 6 linearly 
independent solutions. 
To obtain the characteristic solutions, 
expand all of the terms in 
Equation \ref{eq:xmomentum_dim_linear9}:
\begin{flalign}
    \Big(
 \frac{\partial^6}{\partial{z^6}}
 -i\Big(\frac{2\omega}{\nu}+\frac{\omega}{\kappa}\Big)
 \frac{\partial^4}{\partial{z^4}}
  +\Big(-\frac{\omega^2}{\nu^2}
  -\frac{2\omega^2}{\nu\kappa}
  +\frac{f^2\cos^2\theta{}}{\nu^2}
  +\frac{N^2\sin^2\theta{}}{\nu\kappa}\Big)
  \frac{\partial^2}{\partial{z^2}}
     &  \nonumber &\\
     +i\Big(-\frac{f^2\cos^2\theta{}\omega}{\nu^2\kappa}
     -\frac{N^2\sin^2\theta{}\omega}{\nu^2\kappa}
    +\frac{\omega^3}{\nu^2\kappa}\Big)
    \Big)
    \mathcal{B}
    =i\frac{A\omega{}N^2\sin\theta}{\nu^2\kappa}
    \label{eq:xmomentum_dim_linear13}
\end{flalign}
Therefore our the nonhomogeneous ordinary differential 
equation has the form:
\begin{flalign}
    \Big(
 \frac{\partial^6}{\partial{z^6}}
 +ia_4
 \frac{\partial^4}{\partial{z^4}}
  +a_2
  \frac{\partial^2}{\partial{z^2}}
     +ia_0
    \Big)
    \mathcal{B}
    =if_p
    \label{eq:xmomentum_dim_linear14}
\end{flalign}
where the subscript $p$ denotes ``particular solution'' and
\begin{flalign}
 a_4&=-\Big(\frac{2\omega}{\nu}+
 \frac{\omega}{\kappa}\Big)=-\frac{\omega}{\kappa}\Big(\frac{2}{\Pran}+1\Big)=
 -\frac{\omega}{\kappa}\Big(\frac{2+\Pran}{\Pran}\Big)\nonumber \\
 a_2&=-\frac{\omega^2}{\nu^2}
 -\frac{2\omega^2}{\nu\kappa}
  +\frac{f^2\cos^2\theta{}}{\nu^2}
  +\frac{N^2\sin^2\theta{}}{\nu\kappa}
    =\frac{f^2\cos^2\theta{}+\Pran N^2\sin^2\theta{}
  -\omega^2(1+2\Pran)}{\kappa^2\Pran^2}
  \nonumber \\
 a_0&=\frac{\omega^3}{\nu^2\kappa}-
 \frac{f^2\cos^2\theta{}\omega}{\nu^2\kappa}
     -\frac{N^2\sin^2\theta{}\omega}{\nu^2\kappa}
     =\frac{\omega(\omega^2-f^2\cos^2\theta{}-N^2\sin^2\theta{})}{\kappa^3\Pran}
     \nonumber \\
     f_p&=\frac{A\omega{}N^2\sin\theta}{\nu^2\kappa}
 \label{eq:coeffs}
\end{flalign}
Equation \ref{eq:xmomentum_dim_linear14} has 
the characteristic equation:
\begin{equation}
 \lambda^6
 +ia_4
 \lambda^4
  +a_2
  \lambda^2
    +ia_0
    =0,
    \label{eq:characteristic_equation}
\end{equation}
which has 6 distinct solutions 
for $\lambda$.
The total general solution is the sum of the complementary 
(homogeneous) solutions and the particular (nonhomogeneous) 
solutions:
\begin{equation}
\mathcal{B}(z)=
\mathcal{B}_\textit{C}(z)+
\mathcal{B}_\text{p}(z)
\label{eq:total_solution}
\end{equation}
The complementary solution is therefore of the form:
\begin{equation}
 \mathcal{B}_\textit{C}(z)=c_1\mathrm{e}^{\lambda_1}
 +c_2\mathrm{e}^{\lambda_2}
 +c_3\mathrm{e}^{\lambda_3}
 +c_4\mathrm{e}^{\lambda_4}
 +c_5\mathrm{e}^{\lambda_5}
 +c_6\mathrm{e}^{\lambda_6}
 \label{eq:comp_solution}
\end{equation}
and the particular part of the solution is of the form:
\begin{equation}
 \mathcal{B}_\text{p}=a_p
 \label{eq:part_solution}
\end{equation}
where $a_p$ is an unknown constant.

\subsection{The particular solution}
To solve for $a_p$, substitute the particular solution 
form (Equation \ref{eq:part_solution}
into the nonhomogeneous governing equation 
(Equation \ref{eq:xmomentum_dim_linear14}):
\begin{flalign}
     a_p
    &=\frac{f_p}{a_0} \nonumber \\
    &=\frac{AN^2\sin\theta}
    {\omega^2-f^2\cos^2\theta{}-N^2\sin^2\theta{}}
    \label{eq:part_ode}
\end{flalign}

\subsection*{The complementary solution}
Let $\phi=\lambda^2$ in Equation \ref{eq:characteristic_equation} to obtain:
\begin{equation}
 \phi^3
 +ia_4
 \phi^2
  +a_2
  \phi
    +ia_0
    =0,
   \label{eq:characteristic_equation3}
\end{equation}
where
\begin{equation}
 \lambda_{1,2}=\pm\sqrt{\phi_1}, \qquad 
  \lambda_{3,4}=\pm\sqrt{\phi_2}, \qquad 
   \lambda_{5,6}=\pm\sqrt{\phi_2},
\end{equation}
The solutions to this equation are:
\begin{equation}
 \beta=\sqrt[3]{2 i a_4^3+9 i a_2 a_4+3 \sqrt{3} \sqrt{4 a_2^3+a_4^2 a_2^2+18 a_0 a_4 a_2+4 a_0 a_4^3-27 a_0^2}-27 i a_0}
\end{equation}
\begin{equation}
\phi_1=
\frac{\beta}{3 \sqrt[3]{2}}
-\frac{\sqrt[3]{2} \left(a_4^2+3 a_2\right)}{3\beta}-\frac{i a_4}{3}
\end{equation}
\begin{equation}
\phi_2=
-\frac{\left(1-i \sqrt{3}\right) \beta }{6 \sqrt[3]{2}}  
+\frac{\left(1+i \sqrt{3}\right) \left(a_4^2+3 a_2\right)}{3\cdot2^{2/3} \beta }
-\frac{i a_4}{3}
\end{equation}
\begin{equation}
\phi_3=
-\frac{\left(1+i \sqrt{3}\right) \beta}{6 \sqrt[3]{2}} 
+\frac{\left(1-i \sqrt{3}\right) \left(a_4^2+3 a_2\right)}{3\cdot4^{1/3} \beta }-\frac{i a_4}{3}
\end{equation}

\subsection{Boundary conditions}
For the parameter space we are interested in $c_1=c_3=c_5=0$ 
(to have finite solutions at $z=\infty$): 
\begin{equation}
 \mathcal{B}(z)=
 c_2\mathrm{e}^{-\sqrt{\phi_1}z}
 +c_4\mathrm{e}^{-\sqrt{\phi_2}z}
 +c_6\mathrm{e}^{-\sqrt{\phi_3}z}+a_p
 \label{eq:gen_solution4}
\end{equation}
Now we reinterpret the boundary conditions in terms of $\mathcal{B}$:
\begin{enumerate}
 \item No-slip at the wall ($z=0$) applied to the across slope 
 velocity
 \begin{equation*}
\mathcal{U}=
-\frac{1}{N^2\sin\theta}
\Big(\text{i}\omega
    -\kappa\frac{\partial^2}{\partial{z^2}}\Big)
    i\mathcal{B}=0,
\end{equation*}
leads to the expression:
\begin{equation}
 c_2 (\omega +i \kappa  \phi _1)
 +c_4 (\omega +i \kappa  \phi _2)+
 c_6 (\omega +i \kappa  \phi _3)=-a_p \omega
\end{equation}

 \item No-slip at the wall ($z=0$) applied to the along slope 
 velocity
 \begin{equation*}
    \mathcal{V}=\frac{1}{f\cos\theta}\Big(\frac{1}{N^2\sin\theta}\Big(\omega^2
    +\text{i}\omega(\nu+\kappa)\frac{\partial^2}{\partial{z^2}}-\nu\kappa\frac{\partial^4}{\partial{z^4}}\Big)
    \mathcal{B}-\sin\theta\mathcal{B}-A\Big)=0, 
\end{equation*}
leads to the expression:
\begin{flalign}
& 
  c_6(\text{i}\omega  \phi _3 (\kappa +\nu )-\kappa \nu  \phi _3^2-N^2 \sin ^2\theta+\omega ^2 )+
 \nonumber \\ & \nonumber
 c_2 \left(i \omega  \phi _1 (\kappa +\nu )-\kappa  \nu  \phi _1^2-N^2 \sin ^2\theta 
 +\omega ^2\right)+ \\ &
 c_4 \left(i \omega  \phi _2 (\kappa +\nu )-\kappa  \nu  \phi _2^2-N^2 \sin ^2\theta +\omega ^2\right)
 =AN^2\sin\theta-a_p \left(\omega ^2-N^2 \sin ^2\theta \right)
\end{flalign}

\item The adiabatic wall boundary condition 
 \begin{equation*}
  \frac{\partial\mathcal{B}}{\partial{z}}=0+0i \quad \text{at} \quad z=0
  \label{eq:bc3B}
 \end{equation*} 
 leads to the expression:
\begin{equation}
 -c_4 \sqrt{\phi _2}-c_6 \sqrt{\phi _3}-c_2 \sqrt{\phi _1}=0
\end{equation}

 \end{enumerate} 
Therefore we can solve for the coefficients:
In matrix form:
\begin{equation*}
 \mathbf{E}\cdot\mathbf{x}=\mathbf{y}
\end{equation*}
or
\begin{equation}
\begin{bmatrix} 
 E_{11} & E_{12} & E_{13}  \\
 E_{21} & E_{22} & E_{23}  \\
  E_{31} & E_{32} & E_{33}  
 \end{bmatrix}
 \begin{bmatrix} 
 x_1 \\ 
 x_2 \\ 
 x_3 
 \end{bmatrix}
 =
 \begin{bmatrix} 
 y_1 \\ 
 y_2 \\ 
 y_3 \\ 
 \end{bmatrix}
 \label{eq:coeff_matrix2}
\end{equation} 
where we solve for
\begin{equation}
 x_1=c_2=b_1, \quad
 x_2=c_4=b_2, \quad
 x_3=c_6=b_3,
 \label{eq:bcoeff_def}
\end{equation}
with $\mathbf{E}$ and $\mathbf{y}$ specified the boundary conditions:
\begin{equation*}
 y_1=-a_p\omega{}, \quad
 y_2=AN^2\sin\theta-a_p \left(\omega ^2-N^2 \sin ^2\theta \right), \quad
 y_3=0,
\end{equation*}
\begin{equation*}
 E_{11}=\omega +i \kappa  \phi _1, \quad 
  E_{12}=\omega +i \kappa  \phi _2, \quad 
   E_{13}=\omega +i \kappa  \phi _3, 
\end{equation*}
\begin{equation*}
 E_{21}=i \omega  \phi _1 (\kappa +\nu )-\kappa  \nu  \phi _1^2-N^2 \sin ^2\theta 
 +\omega ^2,
\end{equation*}
\begin{equation*}
 E_{22}=i \omega  \phi _2 (\kappa +\nu )-\kappa  \nu  \phi _2^2-N^2 \sin ^2\theta +\omega ^2,
\end{equation*}
\begin{equation*}
 E_{23}=\text{i}\omega  \phi _3 (\kappa +\nu )-\kappa \nu  \phi _3^2-N^2 \sin ^2\theta+\omega ^2,
\end{equation*}
\begin{equation*}
 E_{31}=-\sqrt{\phi_1}, \quad
  E_{32}=-\sqrt{\phi_2}, \quad
   E_{33}=-\sqrt{\phi_3}, 
\end{equation*}
The solutions for the coefficients in the $\mathcal{B}$ solution (see Equation \ref{eq:bcoeff_def}) are:
\begin{flalign*}
 \Upsilon&=\kappa ^2 \nu  (\sqrt{\phi _2\phi _3} \phi _1+  \sqrt{\phi _1\phi _3} \phi _2 
 +   \sqrt{\phi _1\phi _2} \phi _3) \\ \nonumber
 &+i \kappa  \nu  \omega  (\sqrt{\phi _1\phi _2}
 +  \sqrt{\phi _1\phi _3}
 +  \sqrt{\phi _2\phi _3}+\phi _1+\phi _2+\phi _3)
 \\ \nonumber
 &
 +\nu  \omega ^2+\kappa  N^2 \sin ^2\theta, 
\end{flalign*}
\begin{flalign}
 b_1&=-\frac{1}{\left(\sqrt{\phi _1}-\sqrt{\phi _2}\right) \left(\sqrt{\phi _1}-\sqrt{\phi _3}\right) \Upsilon}
 \big(
 A \kappa  N^2 \sqrt{\phi _2\phi _3} \sin \theta 
 +i A N^2 \omega  \sin\theta 
 +\kappa  N^2 a_p \sqrt{\phi _2\phi _3} \sin ^2\theta \nonumber \\ 
 &+i \kappa  \nu  a_p \omega  (\sqrt{\phi _3} \phi _2^{3/2}
 + \phi _3 \phi _2
 +\sqrt{\phi _2} \phi _3^{3/2})
 +\nu  a_p \omega ^2 \sqrt{\phi _2\phi _3}\big) \label{eq:b1coeff}
\end{flalign}
\begin{flalign}
 b_2&=\frac{1}{\left(\sqrt{\phi _1}-\sqrt{\phi _2}\right) \left(\sqrt{\phi _2}-\sqrt{\phi _3}\right) \Upsilon}
 \Big(
 A \kappa  N^2 \sqrt{\phi _1\phi _3} \sin\theta
 +i A N^2 \omega  \sin\theta 
 +\kappa  N^2 a_p \sqrt{\phi _1\phi _3} \sin ^2\theta  \label{eq:b2coeff} \\
 &
 +i \kappa  \nu  a_p \omega ( \sqrt{\phi _3} \phi _1^{3/2}
 + \phi _3 \phi _1
 + \sqrt{\phi _1} \phi _3^{3/2} )
 +\nu  a_p \omega ^2 \sqrt{\phi _1\phi _3}\Big) 
  \nonumber
\end{flalign}
\begin{flalign}
  b_3&=-\frac{1}{\left(\sqrt{\phi _1}-\sqrt{\phi _3}\right) \left(\sqrt{\phi _2}-\sqrt{\phi _3}\right) \Upsilon}
  \Big(
  A \kappa  N^2 \sqrt{\phi _1\phi _2} \sin\theta
 +iA N^2 \omega  \sin\theta
 +\kappa  N^2 a_p \sqrt{\phi _1\phi _2} \sin ^2\theta  \label{eq:b3coeff} \\ 
 &
 +\text{i}\kappa  \nu  a_p \omega  ( \sqrt{\phi _2} \phi _1^{3/2}
 +  \phi _2 \phi _1
 +  \sqrt{\phi _1} \phi _2^{3/2} )
 +\nu  a_p \omega ^2 \sqrt{\phi _1\phi _2}\Big) 
 \nonumber
\end{flalign}

\subsection{Solutions}
The solutions for the oscillating component of the flow 
(the components with subscript ``O'' in Equations 
\ref{eq:u_decomp} and \ref{eq:b_decomp})
\begin{equation}
b_{\textrm{O},d}(z,t)=\Real\big[
 \mathcal{B}(z)\text{ie}^{\text{i}\omega{t}}\big]=\Real\big[
 \big(c_2\mathrm{e}^{-\sqrt{\phi_1}z}
 +c_4\mathrm{e}^{-\sqrt{\phi_2}z}
 +c_6\mathrm{e}^{-\sqrt{\phi_3}z}+a_p\big)\text{ie}^{\text{i}\omega{t}}\big]
 \label{eq:b_soln}
\end{equation}
where $b_1,b_2,$ and $b_3$ are given by Equations \ref{eq:b1coeff}, \ref{eq:b2coeff}, and \ref{eq:b3coeff}.
The across-slope velocity coefficients are:
\begin{flalign}
 u_1&=b_1(\omega+\text{i}\kappa\phi_1), \label{eq:u1coeff} \\
 u_2&=b_2(\omega+\text{i}\kappa\phi_2), \label{eq:u2coeff} \\
 u_3&=b_3(\omega+\text{i}\kappa\phi_3),\label{eq:u3coeff} 
\end{flalign}
and the along-slope velocity coefficients are:
\begin{flalign}
 v_1&=b_1(\text{i}\omega\phi_1(\kappa+\nu)-\kappa\nu\phi_1^2-N^2\sin^2\theta+\omega^2), \label{eq:v1coeff} \\
  v_2&=b_2(\text{i}\omega\phi_2(\kappa+\nu)-\kappa\nu\phi_2^2-N^2\sin^2\theta+\omega^2), \label{eq:v2coeff} \\
   v_3&=b_3(\text{i}\omega\phi_3(\kappa+\nu)-\kappa\nu\phi_3^2-N^2\sin^2\theta+\omega^2), \label{eq:v3coeff} 
\end{flalign}
and the velocity solutions are
\begin{flalign}
u_{\textrm{O},d}(z,t)&=\Real\big[
 \mathcal{U}(z)\mathrm{e}^{\text{i}\omega{t}}\big] \label{eq:u_soln} \\
 &=\Real\Big[\frac{
 \big(u_1\mathrm{e}^{-\sqrt{\phi_1}z}
 +u_2\mathrm{e}^{-\sqrt{\phi_2}z}
 +u_3\mathrm{e}^{-\sqrt{\phi_3}z}+a_p\omega\big)\mathrm{e}^{\text{i}\omega{t}}}{N^2\sin\theta}\Big]
 \nonumber
\end{flalign}
\begin{flalign}
v_{\textrm{O},d}(z,t)&=\Real\big[
 \mathcal{V}(z)\text{ie}^{\text{i}\omega{t}}\big] \label{eq:v_soln} \\
 &=\Real\Big[\frac{
 \big(v_1\mathrm{e}^{-\sqrt{\phi_1}z}
 +v_2\mathrm{e}^{-\sqrt{\phi_2}z}
 +v_3\mathrm{e}^{-\sqrt{\phi_3}z}+a_p(\omega^2-N^2\sin^2\theta)-AN^2\sin\theta\big)\text{ie}^{\text{i}\omega{t}}}{fN^2\cos\theta\sin\theta}\Big]
 \nonumber
\end{flalign}

\bibliographystyle{jfm}
\bibliography{jfm-instructions}

\end{document}